\documentclass[12pt]{article}
\usepackage{amsfonts}
\usepackage{amssymb}%,letterspace}
\usepackage{graphics,psboxit,amsmath}
\usepackage{subfigure}
\usepackage{graphicx}

\def\hybrid{\topmargin -30pt    \oddsidemargin 0pt %%%%%%%%%%%%%% Archive-30pt
        \headheight 0pt \headsep 0pt
        \textwidth 6.25in       % A4 paper
        \textheight 9.5in       % A4 paper
        \marginparwidth .875in
        \parskip 5pt plus 1pt   \jot = 1.5ex}

\hybrid

\def\baselinestretch{1.2}

\catcode`\@=11

\def\marginnote#1{}
\newcount\hour
\newcount\minute
\newtoks\amorpm
\hour=\time\divide\hour by60
\minute=\time{\multiply\hour by60 \global\advance\minute by-\hour}
\edef\standardtime{{\ifnum\hour<12 \global\amorpm={am}%
        \else\global\amorpm={pm}\advance\hour by-12 \fi
        \ifnum\hour=0 \hour=12 \fi
        \number\hour:\ifnum\minute<10 0\fi\number\minute\the\amorpm}}
\edef\militarytime{\number\hour:\ifnum\minute<10 0\fi\number\minute}
\def\draftlabel#1{{\@bsphack\if@filesw {\let\thepage\relax
   \xdef\@gtempa{\write\@auxout{\string
      \newlabel{#1}{{\@currentlabel}{\thepage}}}}}\@gtempa
   \if@nobreak \ifvmode\nobreak\fi\fi\fi\@esphack}
        \gdef\@eqnlabel{#1}}
\def\@eqnlabel{}
\def\@vacuum{}
\def\draftmarginnote#1{\marginpar{\raggedright\scriptsize\tt#1}}

\def\draft{\oddsidemargin -.5truein
        \def\@oddfoot{\sl preliminary draft \hfil
        \rm\thepage\hfil\sl\today\quad\militarytime}
        \let\@evenfoot\@oddfoot \overfullrule 3pt
        \let\label=\draftlabel
        \let\marginnote=\draftmarginnote
   \def\@eqnnum{(\theequation)\rlap{\kern\marginparsep\tt\@eqnlabel}%
\global\let\@eqnlabel\@vacuum}  }

\def\preprint{\twocolumn\sloppy\flushbottom\parindent 2em
        \leftmargini 2em\leftmarginv .5em\leftmarginvi .5em
        \oddsidemargin -.5in    \evensidemargin -.5in
        \columnsep .4in \footheight 0pt
        \textwidth 10.in        \topmargin  -.4in
        \headheight 12pt \topskip .4in
        \textheight 6.9in \footskip 0pt
        \def\@oddhead{\thepage\hfil\addtocounter{page}{1}\thepage}
        \let\@evenhead\@oddhead \def\@oddfoot{} \def\@evenfoot{} }

\def\numberbysection{\@addtoreset{equation}{section}
        \def\theequation{\thesection.\arabic{equation}}}

\def\underline#1{\relax\ifmmode\@@underline#1\else
        $\@@underline{\hbox{#1}}$\relax\fi}

\def\titlepage{\@restonecolfalse\if@twocolumn\@restonecoltrue\onecolumn
     \else \newpage \fi \thispagestyle{empty}\c@page\z@
        \def\thefootnote{\fnsymbol{footnote}} }

\def\endtitlepage{\if@restonecol\twocolumn \else \newpage \fi
        \def\thefootnote{\arabic{footnote}}
        \setcounter{footnote}{0}}  %\c@footnote\z@ }

\catcode`@=12
\relax

\def\figcap{\section*{Figure Captions\markboth
        {FIGURECAPTIONS}{FIGURECAPTIONS}}\list
        {Figure \arabic{enumi}:\hfill}{\settowidth\labelwidth{Figure
999:}
        \leftmargin\labelwidth
        \advance\leftmargin\labelsep\usecounter{enumi}}}
 \relax
\def\tablecap{\section*{Table Captions\markboth
        {TABLECAPTIONS}{TABLECAPTIONS}}\list
        {Table \arabic{enumi}:\hfill}{\settowidth\labelwidth{Table
999:}
        \leftmargin\labelwidth
        \advance\leftmargin\labelsep\usecounter{enumi}}}
 \relax
\def\reflist{\section*{References\markboth
        {REFLIST}{REFLIST}}\list
        {[\arabic{enumi}]\hfill}{\settowidth\labelwidth{[999]}
        \leftmargin\labelwidth
        \advance\leftmargin\labelsep\usecounter{enumi}}}
 \relax
\makeatletter
\newcounter{pubctr}
\def\publist{\@ifnextchar[{\@publist}{\@@publist}}
\def\@publist[#1]{\list
        {[\arabic{pubctr}]\hfill}{\settowidth\labelwidth{[999]}
        \leftmargin\labelwidth
        \advance\leftmargin\labelsep
        \@nmbrlisttrue\def\@listctr{pubctr}
        \setcounter{pubctr}{#1}\addtocounter{pubctr}{-1}}}
\def\@@publist{\list
        {[\arabic{pubctr}]\hfill}{\settowidth\labelwidth{[999]}
        \leftmargin\labelwidth
        \advance\leftmargin\labelsep
        \@nmbrlisttrue\def\@listctr{pubctr}}}
 \relax
\makeatother
\newskip\humongous \humongous=0pt plus 1000pt minus 1000pt

\newif\ifdtup

\relax

\def\be{\begin{equation}}
\def\ee{\end{equation}}
\def\ba{\begin{eqnarray}}
\def\ea{\end{eqnarray}}

\def\del{\partial}

\def\r{\rho}
\def\a{\alpha}

\def\b{\beta}

\def\g{\gamma}
\def\G{\Gamma}
\def\d{\delta}
\def\D{\Delta}
\def\e{\epsilon}

\def\P{\Pi}

\def\th{\theta}
\def\Th{\Theta}
\def\m{\mu}
\def\n{\nu}
\def\om{\omega}
\def\Om{\Omega}
\def\l{\lambda}
\def\L{\Lambda}
\def\s{\sigma}
\def\S{\Sigma}

\def\cN{{\cal N}}

\def\cR{{\cal R}}

\def\no{\noindent}

\def\qq{\qquad}

\def\IR{\relax{\rm I\kern-.18em R}}

\def \ha {{1\over 2}}

\def \ov {\over}

\def\II{\relax{\rm 1\kern-.35em1}}
\def\IR{\relax{\rm I\kern-.18em R}}
\def\inv{^{\raise.15ex\hbox{${\scriptscriptstyle -}$}\kern-.05em 1}}

\def\cR{{\cal R}}

\def\tL{{\tilde L}}

\begin{document}

\renewcommand{\theequation}{\thesection.\arabic{equation}}
\csname @addtoreset\endcsname{equation}{section}

\newcommand{\beq}{\begin{equation}}
\newcommand{\eeq}[1]{\label{#1}\end{equation}}
\newcommand{\ber}{\begin{eqnarray}}
\newcommand{\eer}[1]{\label{#1}\end{eqnarray}}
\newcommand{\eqn}[1]{(\ref{#1})}
\begin{titlepage}
\begin{center}

JHEP {\bf 0603} (2006) 069 \hfill CERN-PH-TH/2005-192\\
\vskip -.1 cm
\hfill hep--th/0510132\\

\vskip .5in

{\Large \bf Gravity duals for the Coulomb branch \\
of marginally deformed ${\cal N}=4$ Yang-Mills}

\vskip 0.5in

{\bf Rafael Hern\'andez$^1$},\phantom{x} {\bf Konstadinos
Sfetsos}$^2$\phantom{x} and\phantom{x} {\bf Dimitrios Zoakos}$^2$
\vskip 0.1in

${}^1\!$
Theory Division, CERN\\
CH-1211 Geneva 23, Switzerland\\
{\footnotesize{\tt rafael.hernandez@cern.ch}}

\vskip .1in

${}^2\!$
Department of Engineering Sciences, University of Patras\\
26110 Patras, Greece\\
{\footnotesize{\tt sfetsos@des.upatras.gr, dzoakos@upatras.gr}}\\

\end{center}

\vskip .4in

\centerline{\bf Abstract}

\no
Supergravity backgrounds dual to a class of exactly marginal deformations of
${\cal N}=4$ supersymmetric Yang--Mills can be constructed through an $SL(2,\mathbb{R})$
sequence of T-dualities and coordinate shifts. We apply this transformation to
multicenter solutions and derive supergravity backgrounds describing the Coulomb
branch of ${\cal N}=1$ theories at strong 't~Hooft coupling as marginal deformations
of ${\cal N}=4$ Yang--Mills. For concreteness we concentrate to cases with an
$SO(4)\times SO(2)$ symmetry preserved by continuous distributions of D3-branes on
a disc and on a three-dimensional spherical shell. We compute the expectation value of
the Wilson loop operator and confirm the Coulombic behaviour of the heavy quark-antiquark
potential in the conformal case. When the vev is turned on we find situations where a complete
screening of the potential arises, as well as a confining regime where a linear or a
logarithmic potential prevails depending on the ratio of the quark-antiquark separation to
the typical vev scale. The spectra of massless excitations on these backgrounds are
analyzed by turning the associated differential equations into Schr\"odinger problems.
We find explicit solutions taking into account the entire tower of states related
to the reduction of type-IIB supergravity to five dimensions, and hence we go beyond the
$s$-wave approximation that has been considered before for the undeformed case.
Arbitrary values of the deformation parameter give rise to the Heun differential
equation and the related Inozemtsev integrable system, via a non-standard trigonometric
limit as we explicitly demonstrate.

%\vfill
%\vskip .5cm
%\noindent

\end{titlepage}
\vfill
\eject

%\def\baselinestretch{1.2}
%\baselineskip 10 pt
%\noindent

%%%%%%%%%%%%%%%%%%%%%%%%A generalisation of target space%%%%%%%%%%%%%
\def\tT{{\tilde T}}
\def\tg{{\tilde g}}
\def\tL{{\tilde L}}

%%%%%%%%%%%%%%%%%%%%%%%%%%%%%%%%%%%%%%%%%%%%%%%%%%%%%%%%%%%%%%%%%%%%%%
%%%%%%%%%%%%%%%%%%%%%%%%%%%%%%%%%%%%%%%%%%%%%%%%%%%%%%%%%%%%%%%%%%%%%%

\tableofcontents

\def\baselinestretch{1.2}
\baselineskip 20 pt
\noindent

\section{Introduction}

The AdS/CFT correspondence is the first explicit realization of the long suspected
description of the perturbative expansion of an $SU(N)$ gauge theory in the large $N$
limit in terms of a certain string theory \cite{tHooft}. As it relates the weak coupling
regime of ${\cal N}=4$ supersymmetric Yang--Mills to the strong coupling regime of type
IIB string theory compactified on $AdS_5 \times S^5$, and vice versa \cite{Maldacena,GKS,Witten},
a complete formulation of the correspondence would require a precise knowledge
of the strong coupling limit of each theory. However, in spite a detailed proof still remains
a challenge, diverse tests of the correspondence, even beyond the supergravity limit from
the study of sectors with large quantum numbers, have lead to fascinating insights on
both the gauge and gravity sides.

A natural step forward
towards a more complete understanding of the duality is the extension of the correspondence
to less symmetric theories. An appealing candidate is the case of the exactly marginal
deformations of ${\cal N}=4$ Yang-Mills preserving ${\cal N}=1$ supersymmetry \cite{LS},
because they contain a continuous deformation parameter. A gravity dual was recently derived
for the Leigh--Strassler or $\beta$-deformation of the ${\cal N}=4$ theory through an $SL(2,\mathbb{R})$
sequence of T-duality transformations and coordinate shifts \cite{LM}. The construction of this
deformed background has uncovered an interesting set of new tests of the correspondence. A
three-parameter deformation of $AdS_5 \times S^5$ dual to a non-supersymmetric marginal
deformation of ${\cal N}=4$ Yang-Mills was soon after found, as well as a Lax
pair for the bosonic piece of the string in the case of real deformations of the
background, thus implying integrability of the corresponding bosonic string sigma model
\cite{Frolov}. In addition, the energies of semiclassical strings rotating with large angular
momenta in the deformed background were then compared to anomalous dimensions of large gauge
theory operators in the $\beta$-deformed ${\cal N}=4$ theory \cite{FRT}. Diverse features of
the deformed theories and some related spin offs have been further explored recently,
both on the gravity \cite{Gursoy}-\cite{Rashkov} and on the field theory
\cite{NiPre}-\cite{KT}
sides of the correspondence.

In this paper we will extend the construction in \cite{LM} to supergravity backgrounds
describing the Coulomb branch of marginally deformed ${\cal N}=4$ Yang--Mills.
We enter the Coulomb branch of the ${\cal N}=4$ theory when the $SO(6)$ scalar fields
acquire Higgs expectation values. On the gravity side the non-vanishing scalar
expectation values correspond to a multicenter distributions of branes.
We will use some of these backgrounds to find a gravity dual for the Coulomb branch of the
$\beta$-deformation of ${\cal N}=4$ Yang-Mills. The plan of the paper is the following: In
section 2 we will apply a sequence of T-dualities and coordinate shifts to general supergravity
backgrounds with at least a $U(1) \! \times \! U(1)\times \! U(1)$ global symmetry group.
In section 3 we consider two solutions corresponding to two different brane distributions, with
$SO(2) \! \times \! SO(2) \! \times \! SO(2)$ and $SO(4) \! \times \! SO(2)$ global symmetry.
Section 4 is devoted to the issue of supersymmetry for the marginally deformed backgrounds. In
section 5 we probe the geometry of the deformation by computing the expectation value of the
Wilson loop operator along the transversal space for a distribution of D3-branes on a disc
and on a three-dimensional spherical shell preserving $SO(4) \! \times \! SO(2)$ symmetry.
Section 6 contains a detailed analysis of the spectra of massless excitations by studying
the Laplace equation in the backgrounds with $SO(4) \! \times \! SO(2)$ symmetry as well as
the conformal case. The regime of arbitrary values of the deformation parameter leads to
the Heun differential equation, which we relate to a generalized trigonometric limit of
the Inozemtsev integrable system. We conclude in section 7 with a discussion and further
directions of research.

%%%%%%%%%%%%%%%%%%%%%%%%%%%%%%%%%%%%%%%%%%%%%%%%%%%%%%%%%%%%%%%%%%%%%%
%%%%%%%%%%%%%%%%%%%%%%%%%%%%%%%%%%%%%%%%%%%%%%%%%%%%%%%%%%%%%%%%%%%%%%

\section{The general set up}

In this section we will construct supergravity backgrounds dual to the Coulomb
branch of marginally deformed ${\cal N}=4$ supersymmetric Yang-Mills following the
$SL(2,\mathbb{R})$ sequence of T-dualities and coordinate shifts introduced in \cite{LM}.
We will therefore start from supergravity solutions modeling the Coulomb branch of the
${\cal N}=4$ theory at strong coupling. In these backgrounds only a metric and
the self-dual 5-form are turned on. The corresponding expressions can be written
in terms of a harmonic function in $\IR^6$, i.e. the transverse space to the branes.
The metric has the form
\ba
ds_{10}^2 = H^{-1/2} \eta_{\m\n}dx^\m dx^\n + H^{1/2} dx_i dx_i \ ,
\qq \m=0,1,2,3\ ,\quad
i=1,2,\dots , 6\ ,
\label{mutll}
\ea
the self-dual 5-form is given by
\be
F_5 = dA_4+*_{10} dA_4\ ,\qq (dA_4)_{0123i}=-\del_i H^{-1} \ ,
\ee
and the dilaton is a constant $\Phi_0$.
The harmonic function $H$ is in general given by
\be
H=R^4 \int_{\IR^6} d^6 {\bf x'} {\r({\bf x'})\ov |{\bf x}-{\bf x'}|^4 }\ ,
\ee
where the density of the brane distribution is normalized to unity and should be
positive definite. We have already taken the field theory limit so that asymptotically the
space is $AdS_5\times S^5$ with radius $R^4=4 \pi g_{YM}^2 N$ in string units.
The distribution of the brane centers breaks the $SO(6)$ global symmetry of the background.
Generically this breaking is complete, but, in the cases we will be interested, a smaller
subgroup of the ${\cal R}$-symmetry is retained. In particular in order to perform the
combination of dualities in this paper and the same time preserve some supersymmetry, we
will need at least a group isomorphic to $U(1)^3$.

Let us start with a general background in which the ten spacetime coordinates are split into
a seven-dimensional part parametrized by $x^I$, with $I=1,2,\dots , 7$, whereas the remaining three
coordinates form a 3-torus parametrized by the angles $\phi_i$, $i=1,2,3$. We therefore take the
following metric
\be
ds^2_{10} = G_{IJ}(x)dx^I dx^J + \sum_{i=1}^3 z_i(x) d\phi_i^2\ ,
\qq I=1,2,\dots , 7\ ,
\label{met1}
\ee
where, as indicated, the three positive definite functions $z_i$ could depend on the
transverse to the torus coordinates, the $x^I$'s. In the case that the $\phi_i$'s parametrize
the Cartan torus of an undeformed 5-sphere, the $z^i$'s sum up to unity. In all other
cases in which the five-sphere is deformed they do not obey any restriction except of
course those arising from preserving supersymmetry and satisfying the field
equations. The ranges of the angle variables, the associated Killing vectors and their norms are
\ba
\phi_i \in (0,2\pi)\ ,\qq \xi_1=({\bf 0},1,0,0)\ ,\quad \xi_2=({\bf 0},0,1,0)\ ,
\quad \xi_3=({\bf 0},0,0,1)\ ,\quad \xi^2_i=z_i\ ,
\ea
where $i=1,2,3$.
In general there are degeneration surfaces where at least one of the norms vanishes.
In our case they are defined by the equations $z_i=0$. With the above choice for the
ranges of the angular coordinates, regularity of the metric and absence of conical singularities
at the degeneration surfaces requires that the associated ``surface gravity'' (important as a
notion in black hole solutions with Lorentzian signature) equals to one.
In our case this means that
\ba
\kappa^2_i \equiv  {G^{IJ}\del_I \xi_i^2 \del_J \xi^2_i\ov 4 \xi^2_i}\Big |_{z_i=0}=
 {G^{IJ}\del_I z_i \del_J z_i\ov 4 z_i}\Big |_{z_i=0} = 1\ ,
\qq ({\rm no\ sum\ over}\ i=1,2,3) \ .
\label{periio}
\ea

Note that for simplicity we have not allowed mixing terms. Nevertheless such terms
can also be included to allow for more general backgrounds. Under the above assumptions
the self-dual 5-form is given by
\be
F_5 = d C^{(1)} \wedge d\phi_1 \wedge d\phi_2\wedge d\phi_2
+ {1\ov \sqrt{z_1z_2z_3}} *_7 d C^{(1)}\ ,
\ee
for some 1-form $C^{(1)}=C^{(1)}_I dx^I$.
This is clearly self-dual and the 1-form $C^{(1)}$ should be such
that the exterior derivative of the second term is zero.
Hence, there must be a 4-form
$C^{(4)}$ such that
\be
dC^{(4)}= {1\ov \sqrt{z_1z_2z_3}} *_7 dC^{(1)}\ .
\ee
Consider now the change of variables
\be
\phi_1=\varphi_3-\varphi_2\ , \qq \phi_2=\varphi_1+\varphi_2+\varphi_3\ ,\qq
\phi_3=\varphi_3-\varphi_1\ ,
\label{jjg}
\ee
with inverse
\be
\varphi_1={1\ov 3}(\phi_1+\phi_2-2\phi_3)\ ,
\qq \varphi_2={1\ov 3}(\phi_2+\phi_3-2 \phi_1)\ ,
\qq \varphi_3={1\ov 3}(\phi_1+\phi_2+\phi_3)\ .
\label{jkinv}
\ee
This transforms the three-dimensional part of the metric to
\ba
&& \sum_{i=1}^3 z_i d\phi_i^2 = g_{ij} d\varphi_i d\varphi_j
=(z_2+z_3)d\varphi_1^2 + (z_1+z_2)d\varphi_2^2 +
(z_1+z_2+z_3)d\varphi_3^2
\nonumber\\
&& \qq\qq\qq + \ 2 z_2 d\varphi_1 d\varphi_2 + 2 (z_2-z_3)
d\varphi_1 d\varphi_3
+2 (z_2-z_1) d\varphi_2 d\varphi_3\ .
\ea
The $U(1)\times U(1)$ global symmetry is generated by shifts of the angles
$\varphi_1$ and $\varphi_2$.
This shifts correspond to transformations of the three complex superfields
of the $\cN=4$ SYM theory as
\be
(\Phi_1,\Phi_2,\Phi_3)\quad \to\quad
 (\Phi_1,e^{i \a_1} \Phi_2,e^{-i\a_1}\Phi_3)\quad  {\rm and}\quad
(e^{-i\a_2}\Phi_1,e^{i \a_2} \Phi_2,\Phi_3)\ ,
\label{suppo}
\ee
under which the superpotential of the marginally deformed theory
\be
W={\rm Tr}(e^{i\pi\g} \Phi_1 \Phi_2 \Phi_3 - e^{-i\pi\g} \Phi_1 \Phi_3 \Phi_2)\ ,
\ee
remains invariant. The $\cR$-symmetry will then correspond to shifts of the remaining
angle $\varphi_3$. Since all fields transform in the same way under this symmetry
the superpotential is not invariant, and transforms with a weight as it should be.
The supercharges preserved by the original background fit into representations
of the total $U(1)\times U(1) \times U(1)_\cR$ symmetry. The solutions that will be
generated below by the combination of T-dualities and a coordinate shift will preserve
a fraction of the original supersymmetry because they will not involve the angle
$\varphi_3$ and therefore the $U(1)_\cR$ symmetry will not mix with the global
$U(1)\times U(1)$. This will be examined in detail in section 4.

\no
The vacuum structure of the theory (the Coulomb branch) is described by the conditions
\ba
\Phi_1 \Phi_2 = q \Phi_2 \Phi_1\ ,\qq q\equiv e^{-2i\pi \g}\ ,\qq  {\rm and\ cyclic}\ ,
\label{fhjh}
\ea
which are valid for large $N$ (exact for $U(N)$). For general values of $\g$ this requires
traceless $N\times N$ matrices, where in each entry at most one of them is
non-zero.\footnote{Studies on the Coulomb branch of the theory from the gauge theory side
have been done in \cite{Berenstein,Dorey}, where, in addition to the generic behaviour we study here with
gravity duals, it was shown that there exist exceptional Coulomb branches for special values of
the deformation parameter.} We note that in the undeformed theory any traceless$N\times N$
matrix is allowed.

\subsection{The deformation}

We will now deform our background with an $SL(2,\mathbb{R}) \subset SL(3,\mathbb{R})$
transformation of the complete $SL(3,\mathbb{R}) \times SL(2,\mathbb{R})$ duality group
of type-IIB supergravity compactified on the global $U(1) \times U(1)$
torus.\footnote{Prototype examples where such transformations can be performed are provided
by four-dimensional NS-NS backgrounds with two commuting $U(1)$ isometries. Starting with
the exact string background corresponding to the $SU(2)/U(1)\times SL(2,\IR)/U(1)$ coset WZW
models we recognize, after the transformation, the background describing
NS5-branes uniformly distributed on a circle in their transverse space \cite{sfet1}.
A similar case to this was also considered in \cite{LM} where one starts with $\IR^2 \times \IR^2$
in polar coordinates, instead of the coset models.} We repeat here essentially the steps of
\cite{LM} and \cite{Frolov}, but we will no longer consider the conformal constraint $\sum_i z_i=1$.
In order to clarify the derivation we will include all details.

As a first step we will perform a T-duality along the $\varphi_1$ direction. In general,
under T-duality the fields in the NS-NS sector form a closed set and transform among themselves.
Hence for these we may use the standard rules \cite{BUSCHER}. In contrast, the transformation rules
for the R-R sector fields involve those in the NS-NS sector (see, for instance, \cite{BHO}).
We find that the non-zero components of the metric and antisymmetric tensors are
\ba
&&\tilde g_{11}={1\ov z_2+z_3}\ ,
\qq \tilde g_{22}={z_1 z_2+z_1 z_3+z_2 z_3\ov z_2+z_3}\ ,
\nonumber\\
&& \tilde g_{33}=z_1+{4z_2z_3\ov z_2+z_3}\ ,
\qq \tilde g_{23}={2z_2z_3\ov z_2+z_3}-z_1\ ,
\label{tdume}\\
&&
\tilde b_{12}={z_2\ov z_2+z_3} \ ,\qq \tilde b_{13}={z_2-z_3\ov z_2+z_3}\ .
\nonumber
\ea
In addition we obtain for the dilaton
\be
e^{-2\tilde\Phi}= z_2+z_3\ ,
\ee
and for the R-R 3-form
\be
\tilde A^{(3)}_{I\varphi_2\varphi_3}=3 C^{(1)}_I\ .
\ee
We will now perform a coordinate shift
\be
\varphi_2\to \varphi_2+\g \varphi_1\ .
\ee
Then we find for the metric (we use upper case letters to denote the tensors after the shift)
\ba
&& \tilde G_{11}= \tilde g_{11}+ \g^2 \tilde g_{22}=G^{-1} {1\ov z_2+z_3}\ ,
\qq \tilde G_{22}=\tilde g_{22}={z_1 z_2+z_1 z_3+z_2 z_3\ov z_2+z_3}\ ,
\nonumber\\
&& \tilde G_{33}=\tilde g_{33}=z_1+{4z_2z_3\ov z_2+z_3}\ ,
\qq \tilde G_{23}=\tilde g_{23}={2z_2z_3\ov z_2+z_3}-z_1\ ,
\label{tdumdf}\\
&&
\tilde G_{13}= \g \tilde g_{23}=  \g \left({2 z_2z_3\ov z_2+z_3}-z_1\right) \ ,
\qq \tilde G_{12}=\g\tilde g_{22}=\g {z_1 z_2+z_1 z_3+z_2 z_3\ov z_2+z_3}\ ,
\nonumber
\ea
where for notational convenience we have defined
\be
G^{-1}=1+\g^2 (z_1z_2+z_1z_3+z_2z_3)\ .
\label{G}
\ee
The antisymmetric tensor remains unchanged under the coordinate shift, i.e.
$\tilde B_{ab}=\tilde b_{ab}$. For the R-R 3-form this shift produces the non-zero
components
\be
\tilde A^{(3)}_{I\varphi_2\varphi_3}=3 C^{(1)}_I\ ,\qq
\tilde A^{(3)}_{I\varphi_1\varphi_3}=3 \g C^{(1)}_I\ .
\ee
Next we perform again a T-duality transformation along the $\varphi_1$ direction
and find for the metric
\be
G_{ij}=G\left(g_{ij}+ 9  \g^2 z_1 z_2 z_3 \d_{i,3}\d_{j,3}\right) \ ,
\ee
and for the antisymmetric tensor
\ba
&& B_{12}= \g  G (z_1 z_2 + z_1z_3+z_2 z_3)\ ,
\nonumber\\
&& B_{13}=\g G (2 z_2 z_3 -z_1 z_2-z_1 z_3)\ ,
\\
&& B_{23}=\g G (z_1z_3 + z_2z_3- 2 z_1z_2)\ .
\nonumber
\ea
These expressions for the metric and antisymmetric tensor are written in the $\varphi_i$
coordinate system. Returning now, with the help of \eqn{jkinv}, to the original coordinates
we obtain the final result
\be
ds_{10}^2
= G_{IJ}dx^I dx^J + G \left[\sum_{i=1}^3 z_i d\phi_i^2 +  \g^2
z_1z_2z_3 (d\phi_1+d\phi_2+d\phi_3)^2\right]\ .
\label{hfgw}
\ee
For the antisymmetric tensor we get
\be
B= \g G (z_1z_2 d\phi_1\wedge d\phi_2 \ + \ {\rm cyclic\ in\ 1,2,3})\
\ee
and for the dilaton
\ba
e^{2\Phi }= e^{2 \Phi_0}\ G\ .
\label{hjf9}
\ea
Finally the non-vanishing components of the R-R fields are given by
\ba
&& A^{(2)}= -\g C^{(1)}\wedge (d\phi_1 + d\phi_2+ d\phi_3)\ ,
\nonumber\\
\label{RR}
&& A^{(4)}= G C^{(1)} \wedge d\phi_1\wedge d\phi_2\wedge d\phi_3 + C^{(4)}\ .
\ea
The 5-form is then
\be
F_5 = dA^{(4)} - dB\wedge A^{(2)} =
G d C^{(1)} \wedge d\phi_1 \wedge d\phi_2\wedge d\phi_3
+ {1\ov \sqrt{z_1z_2z_3}} *_7 dC^{(1)}\ ,
\label{trht}
\ee
where we have used the identity $dG=-\g^2 [d(Gz_1z_2)+{\rm cyclic\ in\ 1,2,3}]$.
Its self-duality with respect to the deformed metric \eqn{hfgw} can be readily verified.

\no
As usual, the supergravity description is valid if the curvature of the metric remains small
compared to the string scale. Let $R^2$ be the overall scale, in string units, of the metric
\eqn{met1}, which implies that both the $z_i$'s and the seven-dimensional metric scale like $R^2$
and that the combination $\g R^2$ is kept constant. Hence, we require that $R\gg 1$, but in
addition we should make sure that the metric does not degenerate at arbitrary points due to
the combination $ \g R^2$ becoming large. Indeed, if $\g R^2\gg 1$ then we see that the
term $z_i d\phi_i^2$ in the metric \eqn{hfgw} scales like $1/(\g R)^2$ and therefore we should
require that this scale is large compared to unity.\footnote{In the opposite case it is obvious
that curvature invariants, such as the square of the Riemann tensor for the deformed metric
\eqn{hfgw}, will blow up at a generic point of the manifold and the supergravity description will
no longer be valid. Note also that string loop corrections are negligible at a generic point
if $e^{\Phi_0}\ll 1$, independent of the deformation.} To summarize, for the supergravity
description to be valid at a generic point of the manifold we should have
\ba
R\gg 1\qq {\rm and}\qq   \g R^2 \equiv \hat \g \ll R\ .
\label{kfgsu}
\ea
This is the same condition as that obtained for the conformal case in \cite{LM} where it was
also noted that the last condition is sufficient for
the 2-torus parametrized by $\varphi_{1,2}$
to stay much larger than the string scale after the T-dualities.
This can be shown to be the case here as well
by noting that after the combined T-dualities
and the coordinate shift transformation, the volume of the 2-torus is
\be
{\hbox {Vol(2-torus)}}= G (z_1 z_2 +z _1 z_3 + z_2 z_3)^{1/2}
\sim {R^2 \ov 1 + \hat \g^2}\ ,
\ee
where the last expressions indicates, schematically, the way the volume depends on
the parameters $R$ and $\g$ at a generic point. In order for it to be larger than
unity, the constant $\hat \g$ should be smaller than $R$.

\no
The second of the conditions in \eqn{kfgsu} means that $\g\ll 1/R$ for the supergravity
approximation to be valid. Then the deformation parameter
$q$ in \eqn{fhjh} becomes unity. However, this does not mean that we should admit
arbitrary vev distributions to parametrize the Coulomb branch of the theory,
but only those corresponding to scalar fields satisfying conditions \eqn{fhjh}.

\no
We finally note that the periodicities of the angular variables $\phi_i$ remain intact in
the deformed background. To see that, note that the norms of the Killing vectors are now given by
\be
\xi_i^2= G (z_i + \g^2 z_1 z_2 z_3)\ ,\qq i=1,2,3\ .
\ee
A simple computation, using the fact that $z_i=0$ (for at least one $z_i$), shows that
$\del_I \xi_i^2 =\del_I z_i +\cdots$, where the dots indicate terms that vanish at the
degeneration surfaces faster than the indicated first term. Using then the definition
in \eqn{periio},
we find that indeed the ``surface gravity'' equals one, $\kappa^2_i=1$.

%%%%%%%%%%%%%%%%%%%%%%%%%%%%%%%%%%%%%%%%%%%%%%%%%%%%%%%%%%%%%%%%%%%%%%
%%%%%%%%%%%%%%%%%%%%%%%%%%%%%%%%%%%%%%%%%%%%%%%%%%%%%%%%%%%%%%%%%%%%%%

\section{Brane distributions}

In this section we will apply the above construction to derive the explicit form of the
supergravity background describing the Coulomb branch of marginally deformed ${\cal N}=4$
supersymmetric Yang--Mills for two different brane distributions, namely a uniform continuous
distribution of D3-branes on a disc, and on a three-dimensional  spherical shell. These
were first constructed as the extremal limits of rotating D3-brane solutions in \cite{trivedi,sfet1}.
They were also used in several investigations in the literature within the AdS/CFT
correspondence starting with the works of \cite{warn,BS} and belong to the rich class of
examples representing continuous distributions of M- and string theory branes on
higher dimensional ellipsoids \cite{Basfe2,bbs,cglp}.

%%%%%%%%%%%%%%%%%%%%%%%%%%%%%%%%%%%%%%%%%%%%%%%%%%%%%%%%%%%%%%%%%%%%%%
%%%%%%%%%%%%%%%%%%%%%%%%%%%%%%%%%%%%%%%%%%%%%%%%%%%%%%%%%%%%%%%%%%%%%%

\subsection{Solutions with $SO(2) \! \times \! SO(2) \! \times \! SO(2)$ symmetry}

In this case the transverse space coordinates are parametrized as
\ba
\begin{pmatrix} {x_1}\cr{x_2} \end{pmatrix} & = & \sqrt{r^2-b_1} \
\sin\th \ \begin{pmatrix}{\cos\phi_1}\cr{\sin\phi_1} \end{pmatrix} \ ,
\nonumber\\
\begin{pmatrix}{x_3}\cr{x_4} \end{pmatrix} & = & \sqrt{r^2-b_2} \
\cos\th \sin\psi \begin{pmatrix} {\cos\phi_2}\cr{\sin\phi_2} \end{pmatrix} \ ,
\label{jwoi}\\
\begin{pmatrix}{x_5}\cr{x_6} \end{pmatrix} & = &
\sqrt{r^2-b_3} \ \cos\th \cos\psi \begin{pmatrix} {\cos\phi_3}\cr{\sin\phi_3}
\end{pmatrix} \ .
\nonumber
\ea
where $b_i$, $i=1,2,3$ are some real constants.
The ranges of variables are
\be
r\ge {\rm max}(b_1,b_2,b_3)\ ,
\qq 0\le \th,\psi < {\pi\ov 2} \ ,\qq 0\le \phi_{1,2,3} < 2 \pi\ .
\ee
In this coordinate system the metric is given by
\ba
&& ds^2 = H^{-1/2} \eta_{\m\n} dx^\m dx^\n + H^{1/2}{\D r^6\ov f}\ dr^2
\nonumber\\
&& \, + \:\: r^2 H^{1/2}
\Bigg(\Delta_1 d\theta^2 +\Delta_2 \cos^2\theta d\psi^2 +
2 {b_2-b_3\over r^2}\cos\theta\sin\theta\cos\psi\sin\psi d\theta d\psi
\label{dsiib}\\
&& \, + \:\: \Big(1-{b_1\over r^2}\Big) \sin^2\theta d\phi_1^2 +
\Big(1-{b_2\over r^2}\Big)
\cos^2\theta \sin^2\psi d\phi_2^2 +
\Big(1-{b_3\over r^2}\Big) \cos^2\theta\cos^2\psi d\phi_3^2 \Bigg) \ ,
\nonumber
\ea
where the diverse functions are defined as
\ba
H & = & {R^4\ov r^4 \D}\ ,
\nonumber\\
f & = & (r^2-b_1)(r^2-b_2)(r^2-b_3)\ ,
\nonumber\\
\Delta &=& 1 -{b_1\over r^2} \cos^2\theta -{b_2\over r^2}
(\sin^2\theta\sin^2\psi +\cos^2\psi )
- {b_3\over r^2}(\sin^2\theta\cos^2\psi +\sin^2\psi )
\nonumber \\
&+& {b_2b_3\over r^4}\sin^2\theta +{b_1 b_3\over r^4}
\cos^2\theta\sin^2\psi +{b_1b_2\over r^4}\cos^2\theta\cos^2\psi\ ,
\label{d12}\\
\Delta_1
 &=& 1-{b_1\over r^2}\cos^2\theta -
{b_2\over r^2}\sin^2\theta\sin^2\psi -
{b_3\over r^2}\sin^2\theta\cos^2\psi\ ,
\nonumber\\
\Delta_2 &=& 1-{b_2\over r^2}\cos^2\psi -{b_3\over r^2}\sin^2\psi\ .
\nonumber
\ea
This metric can be interpreted as the supersymmetric limit of the most general
non-extremal rotating D3-brane solution \cite{cvetic,RS}, with $b_1$, $b_2$ and $b_3$ related to
the three rotation parameters after an adequate Euclidean continuation.
It was also derived as a domain wall solution within five-dimensional gauged supergravity
and then uplifted to string theory \cite{Basfe2}. We have also taken the field theory
limit so that the space is asymptotically $AdS_5 \times S^5$.

\no
Comparing now \eqn{dsiib} with \eqn{met1} we see that
\ba
z_1 & = & {R^2\ov \D^{1/2}}\left(1-{b_1\ov r^2}\right)\sin^2\th\ ,
\nonumber\\
z_2 & = &
{R^2\ov \D^{1/2}}\left(1-{b_2\ov r^2}\right)\cos^2\th\ \sin^2\psi \ ,
\\
z_3 & = &
{R^2\ov \D^{1/2}}\left(1-{b_3\ov r^2}\right)\cos^2\th\ \cos^2\psi \ .
\nonumber
\ea
The marginally deformed $SO(2) \times SO(2) \times SO(2)$ solution is then obtained by
reinstalling these expressions in \eqn{hfgw}-\eqn{trht}.\footnote{See also
\cite{Ahn} for related work using \eqn{dsiib}.}

\no
Consider now the shift $\varphi_{1,2}\to\varphi_{1,2} + \a_{1,2}$. Then
we easily check that the complex coordinates
\be
w_{1}=x_{1}+ i x_{2}\ ,\qq w_{2}=x_{3} + i x_{4}\ ,\qq w_{3}=x_{5} + i x_{6}\ ,
\ee
transform as the corresponding complex superfields in \eqn{suppo}, and therefore there
is in that respect agreement with the deformed gauge theory.

%%%%%%%%%%%%%%%%%%%%%%%%%%%%%%%%%%%%%%%%%%%%%%%%%%%%%%%%%%%%%%%%%%%%%%%%
%%%%%%%%%%%%%%%%%%%%%%%%%%%%%%%%%%%%%%%%%%%%%%%%%%%%%%%%%%%%%%%%%%%%%%%%

\subsection{Solutions with $SO(4) \! \times \! SO(2)$ symmetry}

These solutions can be obtained by letting  $b_1=-r_0^2$ and $b_2=b_3=0$
into the various general expressions of the previous subsection.
The radial variable obeys $r\ge 0$ and the metric \eqn{dsiib} now becomes
\ba
&& ds^2  = H^{-1/2} \eta_{\m\n} dx^\m dx^\n
+ H^{1/2} {r^2+r_0^2\cos^2\th\ov r^2+r_0^2}\ dr^2
\nonumber \\
&& +  \: \: H^{1/2} \left( (r^2+r_0^2\cos^2\th)
 d\th^2
+ (r^2+r_0^2) \sin^2\th d\phi_1^2
+r^2 \cos^2\th d\Omega_3^2\right) \ ,
\label{ruu1}
\ea
where the harmonic function reduces in this case to
\ba
H =  {R^4\ov r^2 (r^2+r_0^2 \cos^2\th)}\ ,
\label{dj32}
\ea
and the 3-sphere line element is
\ba
d\Om_3^2 = d\psi^2 + \sin^2\psi\ d\phi_2^2 +  \cos^2\psi\ d\phi_3^2\ .
\label{lpj2}
\ea
The forms necessary to compute the NS-NS and R-R field strengths are
\ba
C^{(1)}& = & {R^4 } {r^2+r_0^2\ov r^2+r_0^2 \cos^2\th}\ \cos^4\th\sin\psi
\cos\psi \ d\psi\  ,
\nonumber\\
C^{(4)}& =& -R^{-4}r^2 (r^2+r_0^2\cos^2\th)\
dt \wedge dx_1\wedge dx_2\wedge dx_3 \ .
\ea
In our parametrization the configuration corresponds to a set of D3-branes uniformly
distributed along the $x_1\!-\!x_2$ plane (equivalently for $r=0$ and $\th=\pi/2$)
on a disc of radius $r_0$. In this case we find
\ba
z_1 & = & R^2 \left(1+{r_0^2\ov r^2}\right)
\left(1+{r_0^2 \cos^2\th\ov r^2}\right)^{-1/2}\sin^2\th\ ,
\nonumber\\
z_2 & = &
R^2 \left(1+{r_0^2 \cos^2\th\ov r^2}\right)^{-1/2}\cos^2\th\ \sin^2\psi \ ,
\label{zzzs}
\\
z_3 & = &
R^2 \left(1+{r_0^2 \cos^2\th\ov r^2}\right)^{-1/2}\cos^2\th\ \cos^2\psi \ .
\nonumber
\ea
The marginally deformed $SO(4) \times SO(2)$ background is obtained
once we enter these expressions in \eqn{hfgw}-\eqn{trht}, with the function $G$ given
now by
\be
G^{-1}=1+\hat\g^2 {\cos^2\th\ov r^2+r_0^2\cos^2\th}\left[
(r^2+r_0^2)\sin^2\th + r^2 \cos^2 \th \sin^2\psi \cos^2\psi\right] \ .
\ee

Finally, let us note that for $r_0^2\to -r_0^2$ we get a distribution of
branes on the surface of a four-sphere in the $x_3,\dots , x_6$ space of radius $r_0$
(equivalently for $r=r_0$ and $\th=0$). In that case the radial variable obeys
$r\ge r_0$.

As a general comment we note that the background metric is singular in the location of
the distribution. The reason for this is that in this place the continuum approximation
of the distribution breaks down and it should be replaced by its discrete version. If we
place at each center $N_{\rm center}$ D3-branes such that $1\ll N_{\rm center}\ll N$ then
the gravity description of the gauge theory is still valid with the background corresponding
to $AdS_5\times S^5$ in the $\cN=4$ case, and its exactly marginal deformation in the
$\cN=1$ case. The radius of the space in this case is much smaller that in the continuum
case. It is given by $4\pi g_{\rm YM}^2 N_{\rm center}$ in string units, but nevertheless
it can still be taken to be much larger than the string scale.

In the remaining part of this article we will restrict to the simpler case of
solutions with $SO(4) \times SO(2)$ symmetry, or even to the conformal case. The latter
follows immediately when the vev parameter $r_0=0$. Then the various expressions for the
background fields reduce to those in \cite{LM}.

%%%%%%%%%%%%%%%%%%%%%%%%%%%%%%%%%%%%%%%%%%%%%%%%%%%%%%%%%%%%%%%%%%%%%%%%
%%%%%%%%%%%%%%%%%%%%%%%%%%%%%%%%%%%%%%%%%%%%%%%%%%%%%%%%%%%%%%%%%%%%%%%%

\section{Supersymmetry}

In this section we investigate the issue of supersymmetry for the deformed
supergravity backgrounds and explain in detail the origin of their reduced
supersymmetry.
We start by explicitly showing that the solution for the
Killing spinor in the undeformed case can split into a part which is a singlet of the
$U(1)$ rotations corresponding to the angles $\varphi_1$ and $\varphi_2$, and a part
orthogonal to that. After the T-dualities and the shift only this part survives and remains a Killing
spinor of the deformed theory. For any multicenter metric of the form \eqn{mutll} the Killing
spinor is of the form (this is a mere consequence of the supersymmetry algebra. See, for instance,
\cite{kumar})
\be
\e = H^{-1/8} \e_0\ ,
\ee
where $\e_0$ is a constant spinor subject to the projection
\ba
i \G^{0123}\e_0=\e_0\ ,
\label{hfrp}
\ea
where the indexes refer to the directions along the brane. It is important to realize that
for a different form of the metric such as, in the case of our examples, \eqn{dsiib} and
\eqn{ruu1}, the constant spinor will acquire a coordinate dependence. This can be found by
solving the covariant version of the condition $\del_i \e_0=0$, that is
\ba
\del_i\e_0 + \om_i^{ab}\G_{ab} \e_0 = 0 \ ,\qq \ i=1,2,\dots , 6\ ,
\ea
valid in all coordinate systems. We will concentrate on the background with $SO(4)\times SO(2)$ global
symmetry in the disc case. Then we have the flat metric in $\IR^6$
given by
\ba
ds^2_{\IR^6} & = &{r^2+r_0^2\cos^2\th\ov r^2+r_0^2}\ dr^2
+ (r^2+r_0^2\cos^2\th) d\th^2
+ (r^2+r_0^2) \sin^2\th d\phi_1^2
\nonumber\\
&& +\ r^2 \cos^2\th
(d\psi^2 + \sin^2\psi\ d\phi_2^2 +  \cos^2\psi\ d\phi_3^2) \ ,
\label{rhef}
\ea
as this is read off from \eqn{ruu1}.
Using as a frame basis
\ba
&& e^1= \sqrt{r^2+r_0^2\cos^2\th\ov r^2+r_0^2}\ dr \ ,\:\:\:\: e^2=\sqrt{r^2+r_0^2\cos^2\th}\ d\th \ ,
\:\:\:\: e^3 = r \cos\th\ d\psi \ ,
\nonumber\\
&& e^4= \sqrt{r^2+r_0^2}\sin\th\ d\phi_1\ ,\:\:\:\: e^5= r \cos\th \sin\psi\ d\phi_2 \ ,\:\:\:\:
e^6= r \cos\th \cos\psi\ d\phi_3 \ ,
\ea
we find that the spinor $\e_0$ is given by
\be
\e_0 = e^{\ha f(r,\th)\G_{12}}
e^{{\psi\ov 2}
\G_{13}} e^{\ha \phi_i\s_i}
\bar \e_0\ ,
\ee
where $\bar \e_0$ is a constant spinor and where we have defined
\be
f(r,\th)=\tan\inv \left({r\tan\th \ov (r^2+r_0^2)^{1/2}}\right)\  \:\:
\hbox{and}\ \:\: \s_1=\G_{24}\ ,\quad \s_2=\G_{35}\ ,\quad \s_3= \G_{16}\ .
\label{skpi}
\ee
We are interested in extracting the part of the spinor that is invariant
under variations of the angles $\varphi_1$ and $\varphi_2$. After rewriting the spinor
$\e_0$ in the $\varphi_i$ coordinate system and a simple computation we find that
the required spinor invariant under variations of $\varphi_{1,2}$ is given by
\ba
\e_{0,{\rm inv}}& = &  e^{{1\ov 2} f(r,\th)\G_{12}} e^{{\psi\ov 2} \G_{13}}
e^{\ha [(\s_1+\s_2+\s_3)\varphi_3 + (\s_2-\s_1)\varphi_2 + (\s_2-\s_3)\varphi_1]}
\bar \e_{0,{\rm inv}}\ ,
\nonumber\\
&= & e^{\ha f(r,\th)  \G_{12}} e^{{\psi\ov 2} \G_{13}}
e^{{3\ov 2} \s_3\varphi_3} \bar \e_{0,{\rm inv}}\ .
\label{skpi8}
\ea
The constant spinor $\bar \e_{0,{\rm inv}}$ in terms of $\bar \e_0$ is given by
\be
\bar \e_{0,{\rm inv}}={1\ov 4} (\II-\s_1\s_2-\s_1\s_3-\s_2\s_3)\bar \e_0\ ,
\ee
where the prefactor acts as a projector and by construction we have
\ba
\s_1\bar \e_{0,{\rm inv}}=\s_2 \bar \e_{0,{\rm inv}}= \s_3 \bar\e_{0,{\rm inv}}\ .
\label{projj}
\ea
The part of the spinor that is not invariant under the variations of $\varphi_{1,2}$
will not survive the T-dualities \cite{Bergshoeff,bstsusy,duff} because these are performed
with respect to $\varphi_1$ and $\varphi_2$ as the latter has been shifted by a
$\varphi_1$ component. This conclusion holds within the supergravity approximation,
but in a string theory context we expect to have the broken part of supersymmetry realized
with operators having no field theory analog \cite{bstsusy}.
Related to \eqn{hfrp} and \eqn{projj} projections are expected to arise from a
careful examination of the Killing spinor equations for the deformed background in the generic
case when the vev's are turned on. Being three independent conditions on the spinor we are left
with $\cN=1$ supersymmetry. In the conformal limit \eqn{hfrp} is not necessary, leading to
the $\cN=1$ superconformal case.

%%%%%%%%%%%%%%%%%%%%%%%%%%%%%%%%%%%%%%%%%%%%%%%%%%%%%%%%%%%%%%%%%%%%%%%%%
%%%%%%%%%%%%%%%%%%%%%%%%%%%%%%%%%%%%%%%%%%%%%%%%%%%%%%%%%%%%%%%%%%%%%%%%

\section{Wilson loops and the $q\bar q$-potential}

In this section we will evaluate the Wilson loop operator in the Coulomb branch of
marginally deformed ${\cal N}=4$ supersymmetric Yang--Mills using the $SO(4) \times SO(2)$
background we have described in section 3.2. According to the prescription in
\cite{Rey,maldaloop}, the expectation value of a Wilson loop in the field theory can be
computed by minimizing the Nambu--Goto action for a fundamental string in a given supergravity
background. The Wilson loop is constructed by pulling one brane apart from the brane
distribution, thus giving an expectation value $\vec{\Phi}$ to a Higgs field. The quarks
are the infinitely massive W-bosons connecting the brane distribution to the far away
brane. In order to probe the geometry on the deformed background, we will introduce a
relative angle between the quarks by giving expectation values $\vec{\Phi}_1$ and
$\vec{\Phi}_2$ to two $U(1)$ factors in the global gauge group. In this way we introduce
two relative angles $\vec{\th}_i=\vec{\Phi}_i/|\vec{\Phi}_i|$, and the ends of the Wilson loop,
corresponding to the position of the massive quarks, are located at $r=\infty$ and two
different points $\vec{\th}_1$ and $\vec{\th}_2$ on the transversal space. We will, in
particular, choose coordinates such that the path joining $\vec{\th}_1$ and $\vec{\th}_2$
is parametrized by $\phi_2$ and $\phi_3$ in (\ref{lpj2}). We will take a trajectory with
\be
\th=0\ ,\qq \psi={\pi\ov 4} \ ,
\qq \phi_2=\phi_3\equiv \phi\  , \qq x_{2,3}=\hbox{constant} \ .
\ee
This is consistent with the corresponding equations of motion provided that
the conserved angular momenta for $\phi_2$ and $\phi_3$ coincide.\footnote{We may
consider a trajectory with $\th=\pi/2$ which is also consistent with the equations of
motion. However this choice is not sensitive to the deformation, since the dependence
on the deformation parameter $\hat \g$ drops out. The results
for this trajectory in the particular case of no angular separation for the
quark-antiquark system ($l=0$ below) can be found in \cite{BS}.}
Then, setting these values in (\ref{ruu1})
the reduced four-dimensional metric becomes (in the Euclidean time)
\be
ds^2 = H^{-1/2} \big( d\tau^2 + dx^2 \big) + H^{1/2} \big( dr^2 + r^2 G d\phi^2 \big) \ ,
\ee
with
\be
H = \frac {R^4}{r^2(r^2+r_0^2)} \ ,\qq G^{-1}=1+{\hat \g^2 r^2\ov 4(r^2+r_0^2)}\ .
\ee
We must note that there is also a contribution to this background from the antisymmetric tensor
field. However, as we will only consider static configurations, it will not contribute to the
string action.

%%%%%%%%%%%%%%%%%%%%%%%%%%%%%%%%%%%%%%%%%%%%%%%%%%%%%%%%%%%%%%%%%%%%%%%%
%%%%%%%%%%%%%%%%%%%%%%%%%%%%%%%%%%%%%%%%%%%%%%%%%%%%%%%%%%%%%%%%%%%%%%%%

\subsection{Wilson loops along transversal space}

Before constructing the Wilson loop for the deformed $SO(4) \times SO(2)$ solution,
we will present a discussion which can be used in similar situations when other
more general supergravity backgrounds are considered. In order to confirm with established
notation in the literature, and also emphasize that it has the meaning of energy on the field
theory side, we will use $u$ to denote the radial variable $r$ in all the computations in this
section. The Nambu--Goto action for a static fundamental string stretching along a great
circle in the transversal background is
\ba
S = \frac {T}{2\pi\a'} \int dx \sqrt{ g(u)(\partial_x u)^2 +
f(u)/R^4 + h(u) (\partial_x \phi)^2 } \ ,
\label{SNG}
\ea
where $g(u)= g_{\tau\tau} g_{uu}$, $f(u) = R^4 g_{\tau\tau} g_{xx}$ and
$h(u) = g_{\tau\tau} g_{\phi\phi}$. The factor $T$ comes from the time integration for a static
configuration, because we have taken a rectangular Wilson loop on the boundary, with one side of
length $L$ along the space direction and one of length $T$ along the Euclidean time direction.
Conservation of energy and angular momentum lead to two first order equations
\ba
&& {f\ov \sqrt{g u^{\prime 2} + f/R^4 + h \phi^{\prime 2}}}
=R^2 \sqrt{1-l^2}f_0^{1/2}\ ,
\nonumber\\
&& {h \phi^\prime \ov \sqrt{g u^{\prime 2} + f/R^4 + h \phi^{\prime 2}}}
=l h_0^{1/2}\ ,
\ea
where the two conserved quantities are associated with the constants $u_0$ and $l$, and
the subscript indicates that the corresponding function is computed for $u=u_0$, in which
$u(x)$ develops a minimum. We have used the notation $f_0=f(u_0)$ and $h_0=h(u_0)$,
and the prime denotes derivatives with respect to $x$. Solving these equations for $x$
and $\phi$ in terms of $u$
we find
\be
x = R^2 f_0^{1/2} \sqrt{1-l^2} \int^u_{u_0}du \sqrt{g(u)\ov f(u) F(u)}\ ,
\ee
and
\be
\phi = l h_0^{1/2}  \int^u_{u_0}{du\ov h(u)}
\sqrt{g(u) f(u)\ov F(u)}\ ,
\ee
where we have defined
\be
F(u) \equiv f(u)\left(1-{h_0 l^2\ov h(u)}\right)-(1-l^2)f_0 \ .
\ee
Note that this function vanishes at $u=u_0$, i.e. $F(u_0)=0$. If we place the quark at
$x=L/2$, and the antiquark at $x=-L/2$, the length of the Wilson
loop is
\be
L= 2 R^2 f_0^{1/2} \sqrt{1-l^2} \int^\infty_{u_0}du \sqrt{g(u)\ov f(u) F(u)}\ ,
\ee
and
\be
\D\phi =2 l h_0^{1/2}  \int^{\infty}_{u_0}  {du\ov h(u)}
\sqrt{g(u) f(u)\ov F(u)} \ .
\ee
The total energy of the Wilson loop is divergent because it includes the infinite
contribution from the W-bosons. When we subtract this contribution, the
regularized energy of the quark-antiquark pair reads
\be
E_{q\bar q} = {1\ov \pi} \int^\infty_{u_0} du
\left[\sqrt{g(u) f(u)\ov F(u)} - \sqrt{g(u)}\right]
- {1\ov \pi} \int^{u_0}_{u_{\rm min}} du \sqrt{g(u)} \ ,
\ee
where $u_{\rm min}$ is the minimum value of $u$ allowed by the geometry. In specific examples
we are supposed to solve for the auxiliary parameters $u_0$ and $l$ in terms of the separation
distance $L$ and the separation angle in the internal space $\D \phi$. In practice this cannot
be done explicitly for all values of the energy and length, but in some limited regions instead.
In general, unless $l=0$ we have that $\D\phi\neq 0$. If $l=0$ then $\D\phi=0$ and the angle
$\phi$ remains constant. Then the function $h(u)$ becomes irrelevant since it does not appear
in the expressions for the length and the energy, and all our expressions reduce to those in
\cite{BS}. In the specific examples that follow we will explicitly see that this is indeed the
function encoding the deformation parameter $\hat \g$. Therefore if $l=0$ our results will necessarily
reduce to those for the undeformed theory.

The heavy quark-antiquark potential as computed using the above formulas should obey the
concavity condition
\ba
{dE_{q\bar q}\ov d L}> 0 \ ,\qq {d^2E_{q\bar q}\ov dL^2}\le  0 \ ,
\label{convv}
\ea
stating that the force is always attractive and a non-increasing function of
the separation distance. These conditions were proved quite generally in \cite{Bachas}, and were
investigated in detail for our examples in the case of zero deformation and angular parameters,
i.e. for $\hat \g=l=0$, in \cite{BS}.
For these values of the parameters and for certain trajectories
the potential failed to obey \eqn{convv} for all separation lengths.
The failure is attributed to the fact that the
trajectory approaches the singularity of the D3-brane background corresponding to the location of
the branes and it is precisely this region that gives rise to the violation of the concavity
condition beyond a certain length in the Wilson loop potential. Similar situations will also
arise in this paper for non-zero deformation and angular parameters.

%%%%%%%%%%%%%%%%%%%%%%%%%%%%%%%%%%%%%%%%%%%%%%%%%%%%%%%%%%%%%%%%%%%%%%%%
%%%%%%%%%%%%%%%%%%%%%%%%%%%%%%%%%%%%%%%%%%%%%%%%%%%%%%%%%%%%%%%%%%%%%%%%

\subsection{The conformal limit}

We will now analyze in detail some explicit configurations. In the conformal
case all branes are located at the origin. We have
\be
g(u)=1\ ,\qq f(u)=u^4\ ,\qq h(u)= {u^2 \ov 1+\hat \g^2/4}\ ,
\ee
and $u_{\rm min}=0$. Also the function
\be
F(u)=(u^2-u_0^2) (u^2+u_0^2(1-l^2))\  .
\ee
Then we have for the length (we use 3.137(8 or 12) of \cite{table})\footnote{We will
use the notation ${\bf K}(k)$, ${\bf E}(k)$ and ${\bf \Pi}(n,k)$ for the complete
integrals of the first, second and third kind, respectively.}
\ba
L & = & R^2\sqrt{1-l^2} u_0^2 \int_{u_0^2}^\infty
{d\r\ov \r^{3/2} \sqrt{(\r-u_0^2)(\r+u_0^2(1-l^2))}}
\nonumber\\
& = & {2 R^2\ov u_0\sqrt{(1-l^2)(2-l^2)}}
\left[ (2-l^2) {\bf E} \left(  k \right) - {\bf K}(k) \right] \ ,
\ea
where we have defined $\r=u^2$, and the modulus
\be
k = \sqrt{\frac {1-l^2}{2-l^2}}\  .
\ee
For the angle we find (we use 3.131(8) and 3.137(8) of \cite{table})
\ba
\D \phi  & = & l \sqrt{1+\hat \g^2/4}\ u_0 \int_{u_0^2}^\infty
{d\r \ov \sqrt{\r(\r-u_0^2)(\r+u_0^2(1-l^2))}}
\nonumber\\
& = & 2 \sqrt{1+\hat \g^2/4} {l\ov \sqrt{2-l^2}} {\bf K}(k)\ .
\ea
Therefore we see that the only effect of the deformation is a simple
$\hat \g$-dependent overall factor. The angle is a monotonously increasing
function of the angular parameter $l$ starting from zero. As $l$ increases we reach
its maximum value,
\ba
\D\phi_{\rm max} = \sqrt{1+\hat \g^2/4}\ \pi\ ,\qq {\rm for}\quad  l=1\ .
\label{jfh}
\ea
There are special values of $\hat \g$ for which the string has wound up $n$ times the
circle parameterized by the $\phi$ angle. These are found by setting
$\D\phi_{\rm max}=2 \pi n$ in \eqn{jfh}.
They are given by $\hat \g_n^2 = 4 (4 n^2-1),\ n\in Z$.
Of course the string can wound up for smaller values of $l$ as well. It is not completely
clear to us what the significance of these values for $\g$ is.
For the energy we have (we use 3.141(12) and 3.141(18) of \cite{table})
\ba
E_{q\bar q} & = &
{1\ov 2\pi} \int_{u_0^2}^\infty d\r \left[\sqrt{\r\ov (\r-u_0^2)(\r+u_0^2(1-l^2))}
-{1\ov \sqrt{\r}}
\right] -{u_0\ov \pi}\
\nonumber\\
& = & \ \frac {u_0}{\pi \sqrt{2-l^2}} \left[  {\bf K}(k) -
(2 -l^2) {\bf E}(k)  \right]
\label{potconf}\\
&=& -{2R^2\ov\pi}\ { \left[(2-l^2){\bf E}(k)-{\bf K}(k)\right]^2\ov (2-l^2)\sqrt{1-l^2}}
\ {1\ov L}\ .
\nonumber
\ea
The Coulombic behaviour of the potential is characteristic of cases with
conformal symmetry since the only scale the enters in the various expressions
is the quark-antiquark separation distance. For $l=0$ this becomes
the conformal result for $\cN=4$ Yang-Mills \cite{Rey,maldaloop}.
For larger angular parameter the result is still conformal but with an effective charge
that is a monotonously decreasing function of $l$ until it becomes zero for $l=1$. This
vanishing limit corresponds to a BPS configuration \cite{maldaloop}.

%%%%%%%%%%%%%%%%%%%%%%%%%%%%%%%%%%%%%%%%%%%%%%%%%%%%%%%%%%%%%%%%%%%%%%%%
%%%%%%%%%%%%%%%%%%%%%%%%%%%%%%%%%%%%%%%%%%%%%%%%%%%%%%%%%%%%%%%%%%%%%%%%

\subsection{The disc}

When vev's are turned one we expect that there should be some confining
behaviour for the potential and from the gauge theory side this was studied in \cite{Dorey}.
In the case of a brane distribution along a disc\footnote{Setting $r_0=0$ (no brane
distribution) and $\hat \g=0$ (no marginal deformation) reproduces the situation originally
considered in \cite{maldaloop}. The case $l=\hat \g= 0$ recovers the results in \cite{BS}.}
\be
g(u)=1\ ,\qq f(u)=u^2(u^2+r_0^2)\ ,\qq h(u)= {u^2 (u^2+r_0^2)\ov (1+\hat \g^2/4)u^2 +r_0^2}\ ,
\ee
and $u_{\rm min}=0$. Also the function
\ba
F(u)=(u^2-u_0^2) (u^2+w)\ ,\qq w\equiv (u_0^2+r_0^2)\left(1-{(1+\hat \g^2/4) l^2 u_0^2\ov
(1+\hat \g^2/4)u_0^2+r_0^2}\right)>0\ .
\label{fgrp}
\ea
It turns out that in computing the integrals we have to distinguish between
two cases depending on which one of the two parameters $w$ or $r_0^2$ is larger.
In fact we have that
\be
w>  r_0^2 \qq \Longleftrightarrow \qq u_0^2 >
{(1+\hat \g^2/4) l^2 -1\ov (1+\hat \g^2/4) (1-l^2)} r_0^2\
\ee
and similarly for the reversed inequality. Note that for small enough angular parameter we
always have that $w>r_0^2$, including the smallest value $u_0=0$. It will be also convenient
to use the notation $w_> (w_<)$ to denote the larger (smaller) between the parameters $w$ and
$r_0^2$. Then we have for the length (we use 3.137(8) of \cite{table})
\ba
L & = & R^2\sqrt{1-l^2} u_0 \sqrt{u_0^2+r_0^2} \int_{u_0^2}^\infty
{d\r\ov \r \sqrt{(\r-u_0^2)(\r+r_0^2)(\r+w)}}
\nonumber\\
& = & 2 R^2 \sqrt{1-l^2} \frac {u_0}{w_>} \sqrt{\frac {u_0^2+r_0^2}{u_0^2+w_>}}
\left[ {\bf \Pi} \left( \frac {w_>}{u_0^2 + w_>}, k \right) - {\bf K}(k) \right] \ ,
\label{llgh}
\ea
where we have now defined the modulus
\be
k = \sqrt{\frac {w_>-w_<}{u_0^2+w_>}}\ .
\ee
For the angle we find (we use 3.131(8) and 3.137(8) of \cite{table})
\ba
\D \phi  & = & l h_0^{1/2} \int_{u_0^2}^\infty
d\r
{1+\hat \g^2/4 + r_0^2/\r\ov \sqrt{(\r-u_0^2)(\r+r_0^2)(\r+w)}}
\nonumber\\
& = & \frac {2 l h_0^{1/2}}{\sqrt{u_0^2+w_>}} \left[ \big( 1+ \hat \g^2/4 - r_0^2/w_> \big)
{\bf K}(k)+ \frac {r_0^2}{w_>} {\bf \Pi} \left( \frac {w_>}{u_0^2 + w_>}, k \right) \right] \ .
\label{dnfg}
\ea
Also for the energy we have (we use 3.141(12) and 3.141(18) of \cite{table})
\ba
E_{q\bar q} & = &
{1\ov 2\pi} \int_{u_0^2}^\infty d\r \left[\sqrt{\r+r_0^2\ov (\r-u_0^2)(\r+w)}-{1\ov \sqrt{\r}}
\right] -{u_0\ov \pi}\
\nonumber\\
& = & \ \frac {1}{\pi \sqrt{u_0^2 + w_>}} \left[ (u_0^2 + r_0^2) {\bf K}(k) -
(u_0^2 + w_>) {\bf E}(k)  \right] \ .
\label{enfg}
\ea

\subsubsection{Generic behaviour}

For $u_0\gg r_0$ or, equivalently, at small distances, the vev can be ignored and the
behaviour of the Wilson loop potential turns to be that of the conformal case
\eqn{potconf}.\footnote{We should use the identity ${\bf \Pi}(k^2,k)={\bf E}(k)/(1-k^2)$.} Towards
larger distances the behaviour depends on the relation between the various parameters.
If $(1+\hat \g^2/4)l^2<1$ then the potential becomes monotonically zero at a finite distance.
This is the same as the behaviour of the Wilson loop in the undeformed case corresponding to
$\hat \g=0$ and for zero angular parameter $l=0$ found in \cite{BS}. If on the other hand
$(1+\hat \g^2/4)l^2>1$ the potential, after it turns into positive values, reaches a
maximum and then again approaches zero at a finite distance. In particular, using the
properties of the elliptic functions, we have that as $u_0\to 0$
\ba
L &\simeq &{\pi R^2\ov r_0} \sqrt{1-l^2} \left(1-{u_0\ov r_0}\right)\ ,
\nonumber\\
\D\phi &\simeq &
\pi l \left(1+{\hat \g^2\ov 4} {u_0\ov r_0}\right)\ , \label{u0=0}
\\
E_{q\bar q}& \simeq  & -\left[1-(1+\hat \g^2/4)l^2\right] {u_0^2\ov 4 r_0}\ ,
\nonumber
\ea
Hence we see that the energy goes to zero at a finite value of the length
\be
L_{\rm fin} = {\pi R^2\ov r_0} \sqrt{1-l^2}\ ,
\ee
resulting in a complete screening of charges at a finite distance. If we now use the expression
for the length of the loop in (\ref{u0=0}) to solve for $u_0$ in terms of $L$, we find
the vanishing behaviour
\ba
E_{q\bar q} & \simeq & \! -{1-(1+\hat \g^2/4)l^2\ov 4} \ r_0
\left(L_{\rm fin}-L\ov L_{\rm fin}\right)^2 + \cdots\
\nonumber\\ \label{vanishingE}
& \simeq & \! - {1-(1+\hat \g^2/4)l^2\ov 4(1-l^2)} \ r_0^3
\left(L_{\rm fin}-L\ov \pi R^2\right)^2 + \cdots\ .
\ea
Note that the way the zero energy is approached depends on the sign of
$1-(1+\hat \g^2/4) l^2$, as was mentioned above. For a positive sign the above
length $L_{\rm fin}$ is indeed the maximum distance between the quark and antiquark,
beyond which there is no geodesic connecting them. In this regime the solution is
described by two parallel strings with no interaction potential. The complete screening
behaviour is qualitatively the same as that in \cite{BS}, where the case with
$\hat\g=l=0$ was analyzed in detail. In fact since the complete screening behaviour
happens for $u_0\to 0$, which is when the trajectory approaches the brane distribution,
the continuous approximation breaks down and we should instead use the discrete version of
the brane distribution. Then the potential becomes gradually Coulombic but with $R^2$
in \eqn{potconf} replaced by a much smaller value corresponding to the number of
D3-branes located close to a particular center, i.e. $R^4\to 4 \pi g_{\rm YM}^2 N_{\rm center}$,
with $N_{\rm center }\ll N$. In that sense the screening phenomenon still persists, but it is
just made smoother. This conclusion assumes that $N_{\rm center}\gg 1$, so
that the supergravity description is still valid.

\no
For a negative sign for $1-(1+\hat \g^2/4) l^2$ and depending on the strength of
the parameter $\hat \g$ as compared with $l$, the energy is a single, or a triple-valued
function of the distance (after a critical distance that can be computed numerically).
We have depicted the various behaviours in Figure 1. Beyond the maximum for the energy
the force between the heavy quark and antiquark becomes repulsive and the concavity
condition for the potential \eqn{convv} is not obeyed. This region is not physical and
we think that this behaviour is due to the large value of the deformations parameter
$\hat \g$ in relation to the trajectory approaching the singularity. Then of course
the continuous approximation breaks down and the potential should reach a Coulombic
behaviour as explained above.

%%%%%%%%%%%%%%%%%%%%%%%%%%%%%%%%%%%%%%%%%%%%%%%%
%%%%%%%%%%%%%%%%%%%%%%%%%%%%%%%%%%%%%%%%%%%%%%%%
\begin{figure}[h!]
\begin{center}
\includegraphics[height= 6 cm,angle= 0]{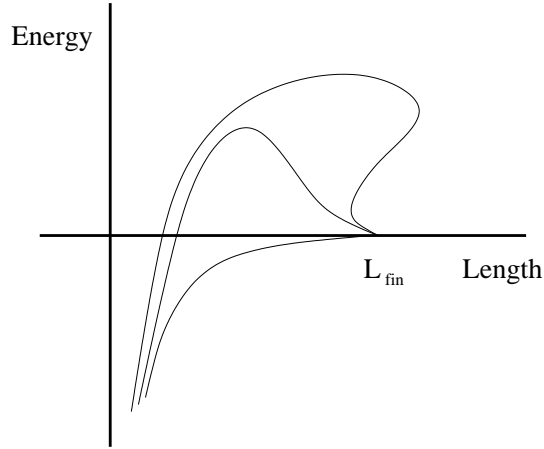}
\end{center}
\caption{The energy \eqn{enfg} as a function of the quark-antiquark distance $L$.
As the value of $\hat{\gamma}$ evolves, the shape of $E_{q\bar{q}}$ is modified and we go
from the lower to the upper curve. For the upper two curves the result is trusted until
the maximum energy is reached. In those cases the maximum length $L_{\rm max}< L_{\rm fin}$.
In the lower curve $L_{\rm max}= L_{\rm fin}$.
}
\label{fig1}
\end{figure}
%%%%%%%%%%%%%%%%%%%%%%%%%%%%%%%%%%%%%%%%%%%%%%%%
%%%%%%%%%%%%%%%%%%%%%%%%%%%%%%%%%%%%%%%%%%%%%%%%

\subsubsection{Confining behaviour}

In order to avoid the apparently unphysical region after the peak of the potential
we should cut off the region beyond this critical point. A careful analysis shows that
this amounts to avoiding the region in the deep IR by essentially restricting the energy
parameter to $u_0\gg r_0/\hat \g$. In the limit of large $\hat \g$, the potential and length
are given by \eqn{llgh} and \eqn{enfg} but with $w=(1-l^2)(u_0^2+r_0^2)$ which arises from
the definition \eqn{fgrp} in the limit $\hat \g \to \infty$. We easily find that the potential
reaches a constant positive value at a finite value of the separation distance, thus again resulting
in a complete screening of charges. They are given by
\ba
E_{q\bar q, {\rm fin}}={r_0\ov \pi}\left[{\bf K}(l)-{\bf E}(l)\right]\ ,
\qq L_{\rm fin} = {\pi R^2\ov r_0}\ .
\ea
The positive value of the maximum energy
depends on the angular parameter $l$ in such a way that it grows when $l$ approaches 1
as $E_{q\bar q, {\rm fin}}\simeq -{r_0\ov 2\pi}\ln(1-l^2)\to \infty$, so that in practice
the potential remains confining.
This allows to consider the limit of a very large deformation parameter and
simultaneously take the angular parameter to one, that is
\ba
\hat \g\gg 1\qq {\rm and } \qq l\to 1\  .
\label{biig}
\ea
This introduces a hierarchy of widely separated scales and in particular note that
\ba
{r_0\ov \hat \g} \ll r_0 \ll {r_0\ov \sqrt{1-l^2}}\ .
\ea
Next we will restrict to the potential corresponding to the corner of the parameter space \eqn{biig}
and further impose the condition
\be
{r_0\ov \hat \g} \ll u_0 \ll {r_0\ov \sqrt{1-l^2}}\ .
\ee
Since the parameter $u_0$ plays essentially the r\^ole of the probe energy,
this means that we will not probe with extremely high energies in the UV,
i.e. extremely higher than the vev value or, equivalently, the separation length will not become
extremely small. This cuts off the conformal region. Similarly, by never probing with
energies in the extreme IR we decouple the region after the maximum of the potential.
We find that in the limit \eqn{biig} the length \eqn{llgh} becomes
\ba
\bar L = {2\ov r_0} {{\bf E}(k)-k'^2 {\bf K}(k)\ov kk'}\ ,
%\frac {2}{r_0^2 u_0} \left[ (u_0^2+r_0^2) {\bf E}(k) - u_0^2{\bf K}(k) \right] \ ,
\label{llgsm}
\ea
where now the modulus simplifies to
\be
k = \sqrt{\frac {r_0^2}{u_0^2+r_0^2}} \ ,\qq \hbox{with } \: k'=\sqrt{1-k^2}\
\ee
and we have defined the finite length
\ba
\qq \bar L={L\ov R^2 \sqrt{1-l^2}} \ .
\label{hfhjw}
\ea
For the angle \eqn{dnfg} we find
\ba
\D \phi =
 \hat \g\ {\bf K}(k)\ ,
\label{dnfsm}
\ea
and for the energy \eqn{enfg} we obtain
\ba
E_{q\bar q}=
%{\sqrt{u_0^2+r_0^2}\ov \pi }\ \Big( {\bf K}(k) - {\bf E}(k)  \Big) \ .
{r_0\ov \pi}{{\bf K}(k)-{\bf E}(k)\ov k}\ .
\label{enfsmg}
\ea
The potential obeys now the concavity conditions \eqn{convv}, and its behaviour is
depicted in Figure 2.

%%%%%%%%%%%%%%%%%%%%%%%%%%%%%%%%%%%%%%%%%%%%%%%%
%%%%%%%%%%%%%%%%%%%%%%%%%%%%%%%%%%%%%%%%%%%%%%%%
\begin{figure}[h!]
\begin{center}
\includegraphics[height= 5.5 cm,angle= 0]{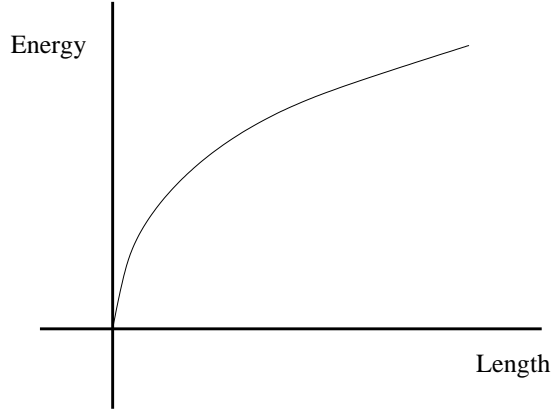}
\end{center}
\caption{The energy \eqn{enfsmg} as a function of the quark-antiquark separation.
For large values of the distance we find the logarithmic confining dependence \eqn{logr}.
For smaller separations the behaviour is linear, \eqn{liin}.}
\label{fig2}
\end{figure}
%%%%%%%%%%%%%%%%%%%%%%%%%%%%%%%%%%%%%%%%%%%%%%%%
%%%%%%%%%%%%%%%%%%%%%%%%%%%%%%%%%%%%%%%%%%%%%%%%

Consider the large values
\be
u_0\gg r_0\ \: ({\rm and}\ u_0\ll {r_0\ov \sqrt{1-l^2}})
\quad \Leftrightarrow \quad \bar L\ll {1\ov r_0}
\ \: ({\rm and}\ \bar L\gg {\sqrt{1-l^2}\ov r_0})\ ,
\ee
where the inequalities inside the parentheses emphasize that energies cannot become extremely
large and correspondingly the lengths cannot be too short. Otherwise the limit \eqn{biig}
that resulted into \eqn{llgsm}-\eqn{enfsmg} is not self-consistent in a mathematical
and a physical sense. Then we obtain for the energy a linear behaviour as a function of the
separation distance,
\ba
E_{q\bar q}\simeq {r_0^2\ov 4 u_0}\simeq {r_0^2\ov 2\pi} \bar L\to 0\ .
\label{liin}
\ea
For the angle we obtain the large limiting value
\be
\D\phi = \hat \g\ {\pi\ov 2}\ .
\ee
Let us now consider the opposite limit of low values
\be
u_0\ll r_0\ ({\rm and}\ u_0\gg {r_0\ov \hat \g})\quad \Leftrightarrow \quad
\bar L\gg {1\ov r_0}
\ ({\rm and}\ \bar L\ll {\hat \g\ov r_0})\ ,
\ee
where, similarly to the previous case, the inequalities in the parentheses imply that
the region in the deep IR is avoided for the limit \eqn{biig} to be self-consistent.
For the energy we find
\ba
E_{q\bar q}\simeq -{r_0\ov \pi}\ln\left( u_0\ov r_0\right)
\simeq {r_0\ov \pi}\ln\left(r_0\bar L\right)\ \to \infty\ .
\label{logr}
\ea
Hence we have a confining behaviour for the heavy quark-antiquark potential,
although the dependence on the separation distance is not linear but logarithmic.
For the angle we obtain again a logarithmic behaviour
\be
\D\phi \simeq \hat \g \ \ln (r_0 \bar L)\ .
\ee
It is worth commenting that a logarithmic form for a confining potential instead of a linear
one is not strange to high energy physics. Long ago it was shown that a logarithmic potential
is unique in producing spectra with energy level spacing independent of the
particle mass \cite{quigg}, an idea that arose from related observations at particle spectra at the time.
In the same work the logarithmic potential was used in relation to quarkornium spectra
for various families of heavy $q\bar q$ bound states (for related reviews on the use of
Schr\"odinger equation in heavy quark systems see \cite{QuiggREV}). This phenomenological
way of introducing a logarithmic potential in Particle Physics had no theoretical support within
perturbative approaches to QCD. More recently, in the context of potential non-relativistic QCD,
logarithmic terms arose, in calculations valid for length scales less than $1/\L_{\rm QCD}$,
as modifications of the effective charge in the Coulombic
potential (see \cite{brambilla} and references therein).
Given our results, it will be interesting to investigate
the issue of a logarithmic confining behaviour from a gauge theoretical view point.

%%%%%%%%%%%%%%%%%%%%%%%%%%%%%%%%%%%%%%%%%%%%%%%%%%%%%%%%%%%%%%%%%%%%%%%%
%%%%%%%%%%%%%%%%%%%%%%%%%%%%%%%%%%%%%%%%%%%%%%%%%%%%%%%%%%%%%%%%%%%%%%%%

\subsection{The sphere}

This case follows from the expressions above if we let $r_0^2\to -r_0^2$ and
take into account that now $u_{\rm min}=r_0$. Therefore we have
\be
g(u)=1\ ,\qq f(u)=u^2(u^2-r_0^2)\ ,\qq h(u)= {u^2 (u^2-r_0^2)\ov (1+\hat \g^2/4)u^2 -r_0^2}\ .
\ee
Also the function
\be
F(u)=(u^2-u_0^2) (u^2+w)\ ,\qq w\equiv (u_0^2-r_0^2)\left(1-{(1+\hat \g^2/4) l^2 u_0^2\ov
(1+\hat \g^2/4)u_0^2-r_0^2}\right)\ .
\ee
It turns out that in this case we always have, although $w$ is not a strictly positive
constant, that $w>-r_0^2$, and therefore we do not have to distinguish between two different
cases depending on the values of $w$ and $r_0^2$, as in the disc case.
%In fact we have that
%\be
%w>  -r_0^2 \qq \Longleftrightarrow \qq u_0^2 >
%{1-(1+\hat \g/4) l^2 \ov (1+\hat \g/4) (1-l^2)} r_0^2\
%\ee
%and similarly for the reversed inequality.

Then we have for the length (we use 3.137(8) of \cite{table})
\ba
L & = & R^2\sqrt{1-l^2} u_0 \sqrt{u_0^2-r_0^2} \int_{u_0^2}^\infty
{d\r\ov \r \sqrt{(\r-u_0^2)(\r-r_0^2)(\r+w)}}
\nonumber\\
& = & 2 R^2 \sqrt{1-l^2} \frac {u_0}{w} \sqrt{\frac {u_0^2-r_0^2}{u_0^2+w}}
\left[ {\bf \Pi} \left( \frac {w}{u_0^2 + w}, k \right) - {\bf K}(k) \right] \ ,
\ea
where we have defined $\r=u^2$, and the modulus
\be
k = \sqrt{\frac {w+r_0^2}{w+u_0^2}}\ .
\ee
For the angle we find (we use 3.131(8) of \cite{table})
\ba
\D \phi  & = & l h_0^{1/2} \int_{u_0^2}^\infty
d\r
{1+\hat \g^2/4 - r_0^2/\r\ov \sqrt{(\r-u_0^2)(\r-r_0^2)(\r+w)}}
\nonumber\\
& = & \frac {2 l h_0^{1/2}}{\sqrt{u_0^2+w}} \left[ \big( 1+ \hat \g^2/4 + r_0^2/w \big)
{\bf K}(k)- \frac {r_0^2}{w} {\bf \Pi} \left( \frac {w}{u_0^2 + w}, k \right) \right] \ .
\ea
For the regularized energy we now have (we use 3.141(12) of \cite{table})
\ba
E_{q\bar q} & = &
{1\ov 2\pi} \int_{u_0^2}^\infty d\r \left[\sqrt{\r-r_0^2\ov (\r-u_0^2)(\r+w)}-{1\ov \sqrt{\r}}
\right] -{u_0-r_0\ov \pi}\
\nonumber\\
& = & \ \frac {1}{\pi \sqrt{u_0^2 + w}} \left[ (u_0^2 - r_0^2) {\bf K}(k) -
(u_0^2 + w) {\bf E}(k)  \right] + {r_0\ov \pi}\ .
\ea
As before, for short distances we may ignore the vev and the behaviour of the Wilson loop remains
that of the conformal case. For larger values of $r_0$ we may easily show, using the properties of
the elliptic functions, that the length goes again to zero and therefore the energy is a double
valued function of the separation length. As before, if we consider the limit of large $\hat\g$
together with $l\to 1$ we will cut off the conformal part of the potential. In these limiting
case we obtain for the length and the energy the simplified expressions
\ba
\bar L={2\ov r_0} {k'\ov k} [{\bf K}(k)-{\bf E}(k)]\
\ea
and
\ba
E_{q\bar q}={r_0\ov \pi}\left({k'^2\ov k} {\bf K}(k)-{1\ov k}{\bf E}(k) +1\right)\ ,
\ea
where now the modulus is $k=r_0/u_0$ and $\bar L$ is defined as in \eqn{hfhjw}.
Using properties of the elliptic functions one easily verifies that the separation
$L$ goes to zero both for $k\to 0$ and $k\to 1$ and therefore there is no confining behaviour.
We note that there is such behaviour (linear) for the $\th={\pi\ov 2}$ trajectory
which however, as remarked in footnote 5, is not sensitive to the deformation.
The energy is a doubled valued function of the length. The angular distance now becomes
\ba
\D \phi = \hat \g k' {\bf K}(k)\ .
\ea
%\ba
%L &\simeq &
%\nonumber\\
%\D\phi &\simeq &
%\\
%E_{q\bar q}& \simeq  & \ ,
%\nonumber
%\ea

%%%%%%%%%%%%%%%%%%%%%%%%%%%%%%%%%%%%%%%%%%%%%%%%%%%%%%%%%%%%%%%%%%%%%%%%%%%%%%%%
%%%%%%%%%%%%%%%%%%%%%%%%%%%%%%%%%%%%%%%%%%%%%%%%%%%%%%%%%%%%%%%%%%%%%%%%%%%%%%%%

\section{The wave equation}

In this section we study the massless scalar field equation
\be
\Box \Psi = {1\ov \sqrt{-G}} e^{-2\Phi}\del_{M} \sqrt{-G} e^{-2 \Phi}
G^{MN} \del_N \Psi = 0 \ ,
\ee
in the deformed $SO(4) \times SO(2)$ background. Quite
generally, for the background metric \eqn{hfgw} and the dilaton \eqn{hjf9}, it
can be shown that the above equation can be written as a deformation of the
corresponding equation for the undeformed background,
\ba
\Box \Psi = \Box_{\g=0} \Psi +\g^2 \l_{ij}\ \del_i \del_j \Psi  = 0\  \ ,
\label{jedhg}
\ea
where we have defined
\be
\l_{ij}\equiv  {z_1z_2 +z_1z_3 +z_2 z_3\ov z_i}\d_{ij}
-{z_1 z_2 z_3 \ov z_i z_j}\ .
\ee
Let us now study each of the pieces entering this differential equation.

%%%%%%%%%%%%%%%%%%%%%%%%%%%%%%%%%%%%%%%%%%%%%%%%%%%%%%%%%%%%%%%%%%%%%%%%%%%%%%%%
%%%%%%%%%%%%%%%%%%%%%%%%%%%%%%%%%%%%%%%%%%%%%%%%%%%%%%%%%%%%%%%%%%%%%%%%%%%%%%%%

\subsection{The undeformed case}

First we will examine the Laplacian for the undeformed case with $\hat \g=0$.
This case is on its own
already very interesting because it describes the spectrum of dilaton, transverse graviton
and gauge field fluctuations in the undeformed background.\footnote{These spectra are in
fact degenerate and the corresponding fluctuations are interrelated and can be written in terms
of the dilaton fluctuations. These have been exposed in a series of papers,
for the dilaton-gravity in \cite{BS2} and for the dilaton-gauge fields in \cite{bianchi,BS3}.
This interrelation is essentially due to the fact that all fluctuations correspond to
fields that belong to the same $\cN=4$ supergravity multiplet \cite{bianchi}. It would be very
interesting to see whether such a coincidence of spectra persists in the marginally deformed
$\cN=1$ case as well.} Within the AdS/CFT correspondence these fluctuations
correspond to gauge theory operators. In that respect the $s$-wave case has been explicitly first
solved in \cite{warn,BS}. In that case the description can be made also using the five-dimensional gauged
supergravity arising from reduction of type-IIB supergravity on $S^5$. Here, we will see that
we can go beyond gauged supergravity and explicitly solve this equation in all generality.
We will shall see below that, turning on the deformation will not
affect dramatically most of our analysis of this subsection.

For the background corresponding to the disc distribution we make the ansatz
\ba
\Psi= {1\ov (2\pi)^{2}} e^{ik\cdot x}  \Psi_\perp(r,\th,\phi_1,\psi,\phi_2,\phi_3)\ .
\label{ans1}
\ea
Obtaining the corresponding differential equations below for the sphere or the conformal
cases simply amounts to analytically continuing the parameter $r_0^2$ to $-r_0^2$ or setting it
to zero, respectively. Then after some algebra we find the following second order differential
equation for the Laplace operator acting on the amplitude
\ba
\Box_{\g = 0} \Psi \!\! & \sim & \!\! {r\ov R^2 \sqrt{r^2+r_0^2\cos^2 \th}}
\Bigg[ {R^4M^2\ov r^2} +{1\ov r^3 } \del_r r^3 (r^2+r_0^2) \del_r
+ {1\ov \sin\th \cos^3\th}\del_\th \sin\th \cos^3\th \del_\th
\nonumber\\
& + & \!\! \left({1\ov \sin^2\th} -{r_0^2\ov r^2+r_0^2}\right) \del^2_{\phi_1}
+ \left({1\ov \cos^2\th}+{r_0^2\ov r^2}\right) \D_{S^3} \Bigg]\Psi_\perp \ ,
\label{hfdjew}
\ea
where the mass eigenvalue is defined as
\be
M^2\equiv -k^2 \ ,
\ee
and where we have omitted an overall coefficient and the factor from the exponential.
Obviously the Laplace equation corresponding to the operator in \eqn{hfdjew} admits
solutions via the separation of variables method. Let
\be
\Psi_\perp(r,\th,\phi_1,\psi,\phi_2,\phi_3) = {1\ov \sqrt{2\pi}} e^{i n\phi_1}
\Psi_{S^3}(\psi,\phi_2,\phi_3) \psi^{(1)}(\th)\psi^{(2)}(r)\ , \qq n\in {\mathbb Z} \ .
\ee
For the $\Psi_{S^3}$ piece we get the standard eigenvalue equation on $S^3$
\ba
\D_{S^3} \Psi_{S^3} = - l(l+2)\Psi_{S^3}\ ,\qq l=0,1,\dots\ ,
\label{sph3}
\ea
or, explicitly,
\ba
{1\ov \sin 2\psi}{\del\ov \del\psi}\left(\sin 2\psi {\del\Psi_{S^3} \ov \del\psi}\right)
+ \left(l(l+2) + {\del_{\phi_2}^2\ov \cos^2\psi}
+{\del_{\phi_3}^2\ov \sin^2\psi}\right)\Psi_{S^3}=0\ .
\label{diejr1}
\ea
The normalized solution is given by
\ba
\Psi_{S^3,l,n_2,n_3}(\th,\phi_2,\phi_3)
& = &   A_{l,n_1,n_2} e^{i(n_2\phi_2+n_3\phi_3)}(\sin{\psi})^{|n_3|}(\cos{\psi})^{|n_2|}
P^{(|n_3|,|n_2|)}_{k}(\cos 2 \psi)\ ,
\nonumber\\
&&  {l}-{|n_2|}-{|n_3|}=2 k \ ,\qq k=0,1,\dots \ ,
\label{doo3}
\ea
where the normalization constant is
\ba
A^2_{l,n_1,n_2} = {l+1\ov \pi} {\G(\ha l+\ha |n_2|+\ha |n_3|+1)
\G(\ha l-\ha |n_2|-\ha |n_3|+1)\ov \G(\ha l+\ha |n_3|-\ha |n_2|+1)
\G(\ha l-\ha |n_3|+\ha |n_2|+1)}\ .
\label{noorm3}
\ea
The mass eigenvalue has degeneracy
\ba
d_{3,l}& = & \sum_{k=0}^\infty \sum_{n_2,n_3=-\infty}^\infty \d_{2 k + |n_2|+|n_3|,l}
 =  \sum_{k,p=0}^\infty \d_{2k+p,l}\sum_{n_2,n_3=-\infty}^\infty \d_{|n_2|+|n_3|,p}
\nonumber\\
& = &
\sum_{k,p=0}^\infty (4 p+\d_{p,0}) \d_{2k+p,l}
\label{deg1}\\
& = & (l+1)^2\ ,
\nonumber
\ea
which is the correct expression.\footnote{
A general eigenstate of the Laplace operator on the unit $n$-sphere has energy eigenvalue
and degeneracy given by (see, for instance, \cite{vN})
\ba
&&E_{n,j}= j(j+n-1)\ ,\qq j =0,1,\dots \ ,
\nonumber \\
&& {d}_{n,j} = {(2 j +n-1) (j+n-2)!\ov (n-1)! j!}\ .
\label{made}
\ea
}
Using the above we obtain, for the other factors,
\ba
{1\ov \sin\th \cos^3\th}{d\ov d\th}\sin\th \cos^3\th {d\psi^{(1)}\ov d\th}
+ \left(E-{n^2\ov \sin^2\th} -{l(l+2)\ov \cos^2 \th}\right)\psi^{(1)} = 0 \
\label{gfhr1}
\ea
and
\ba
{d\ov dr } r^3 (r^2+r_0^2) {d\psi^{(2)}\ov dr} +  \left[ (M^2 R^4-l(l+2)r_0^2) r
+ {r_0^2 r^3\ov r^2+r_0^2} n^2 - E r^3\right]\psi^{(2)} =0 \ ,
\label{gdh4}
\ea
where $E$ is the separation of variables constant. The measure for the resulting
Hilbert space is
\ba
%\sqrt{-G} G^{tt} =d^4 x dr d\th \sin\th\cos^3\th \ .
\sqrt{-G} G^{tt} =d^4 x (dr r) (d\th  \sin\th\cos^3\th)(d\psi \sin\psi \cos\psi) \ .
\label{meas1}
\ea
It is quite remarkable that the equations for the angular part, and in particular
\eqn{gfhr1}, do not depend on the parameter $r_0$ characterizing the vev of the scalar
fields. Hence we expect that $E$ should be quantized and written as $E=j(j+4)$, $j=0,1,\dots $
as appropriate for the Laplacian on the undeformed $S^5$. This is not immediately
apparent, but it works precisely like that as we shall prove. Writing
\ba
\psi^{(1)} = (1-x)^{|n|/2} (1+x)^{l/2} F(x)\ ,\qq x=\cos2\th\ ,\quad |x|\leq 1\ ,
\label{hfg2}
\ea
we obtain an equation for $F(x)$
%\be
%(1-x^2)F^{\prime\prime}+ \left[l+1-n-(l+n+3)x ]F^\prime
%+[E/4 -l(l+2n +4)/4 - n (1+n/4)]F = 0 \ .
%\ee
which is the Jacobi differential equation with $\a=|n|$ and $\b=l+1$ in the standard
notation.\footnote{More precisely,
if we change variable as $z=(1+\cos 2\th)/2$,
then the differential equation for $\psi^{(1)}$ transforms into a second order
ordinary differential equation of the Fuchsian type with three regular singularities
at $z=0,1$ and $\infty$. Such an equation can be transformed into the canonical
form of a hypergeometric equation. The latter is transformed to the Jacobi equation
by a simple variable change $x=2 z-1$($=\cos 2\th$ as in \eqn{hfg2}) which has
polynomial solutions provided the parameters $a,b,c$ in the standard notation are
related to, at least, an integer. The same reasoning applies for the differential
equation for $\psi^{(2)}$ that is considered separately for the conformal, the disc
and the sphere cases below. We explain in section 6.3 that in the deformed case
the nature of the singularity at $z=\infty$ changes from a regular to an irregular one
and as a result we cannot solve the corresponding differential equation by elementary methods.}
It has the Jacobi polynomials as a complete set of orthonormal solutions
provided that the parameter $E$ is quantized as
\ba
E_{m,l,n}=(l+|n|+2m)(l+|n|+2 m +4)\ ,\qq m= 0,1,\dots\ .
\label{jhgf}
\ea
Indeed this takes the form $j(j+4)$, with $j=l+|n|+2 m$, as advertised. Moreover, the
degeneracy of this state works out as it should be. Indeed, note that in general
the degeneracy is given by
\ba
d_{5,j}&=&\sum_{m,l=0}^\infty \sum_{n=-\infty}^\infty
d_{3,l} \d_{l+|n|+2 m,j} =\sum_{m,p=0}^\infty  \d_{p+2 m,j}
\left[\sum_{l=0}^\infty \sum_{n=-\infty}^\infty (l+1)^2
\d_{l+|n|,p}\right]
\nonumber\\
& = & {1\ov 3} \sum_{m,p=0}^\infty (p+1)(2 p^2 + 4 p+3)\d_{p+2 m,j}
\label{deg22}\\
& = & {1\ov 12} (j+1) (j+2)^2 (j+3)\ ,
\nonumber
\ea
which is the correct expression (see \eqn{made} with $n=5$). Then
\ba
\psi^{(1)}_{m,l,n}(\th)= B_{m,l,n} \sin^{|n|}\th \cos^l\th P^{(|n|,l+1)}_m(\cos 2\th) \ ,
\label{fgjw}
\ea
where the normalization constant obeys
\be
B_{m,l,n}^2 = 2(2m+2+l+|n|) {\G(m+1) \G(m+l+|n|+2)\ov \G(m+|n|+1) \G(m+l+2)}\ .
\ee
The integration measure that determined the normalization constant is $d\th \sin\th
\cos^3\th$, with $0\le \th\le \pi/2$.

\no

Let us now present the solution to the remaining radial equation. We will have to
distinguish between the conformal, the disc and the sphere cases.

%%%%%%%%%%%%%%%%%%%%%%%%%%%%%%%%%%%%%%%%%%%%%%%%%%%%%%%%%%%%%%%%%%%%%%%%
%%%%%%%%%%%%%%%%%%%%%%%%%%%%%%%%%%%%%%%%%%%%%%%%%%%%%%%%%%%%%%%%%%%%%%%%

\subsubsection{The conformal limit}

In this case we have to solve equation \eqn{gdh4} after setting the parameter $r_0=0$. This
can be written as a Schr\"odinger equation by letting
\be
\psi^{(2)}=z^{3/2} \Psi(z)\ ,\qq  z = {1\ov r}\ ,\quad
z\ge 0\ .
\ee
Then $\Psi$ satisfies the standard Schr\"odinger equation with potential
\ba
V(z)= {15/4+E_{m,l,n}\ov  z^2}
\label{pot1con}
\ea
and eigenvalue $M^2R^4$. This potential is positive definite and vanishes for large values
of $z$. Hence we expect a continuous spectrum with no mass gap. The explicit solution is
\ba
\psi^{(2)}= \sqrt{M} z^2 J_{j+2}(Mz)\ ,\qq j=0,1,\dots\ ,
%\left\{ \begin{array}{l l} J_2(Mz) \ ,& j=0\ ,\\
%J_{i\m}(M z)\  ,& \m^2=j^2+4 j-4\ ,\quad j=1,2,\dots \end{array}\right\}\ ,
\label{k1wsft}
\ea
where $j$ is the integer that parametrizes the eigenvalue in \eqn{jhgf} as we have
mentioned above.

%%%%%%%%%%%%%%%%%%%%%%%%%%%%%%%%%%%%%%%%%%%%%%%%%%%%%%%%%%%%%%%%%%%%%%%%
%%%%%%%%%%%%%%%%%%%%%%%%%%%%%%%%%%%%%%%%%%%%%%%%%%%%%%%%%%%%%%%%%%%%%%%%

\subsubsection{The disc}

In the previous section we have found a vanishing behaviour beyond a maximal length
for the interaction potential between a heavy quark and an antiquark.
A maximal length implies the existence
of a set of small energy scales that we can not reach. Therefore the gauge theory must
generate a mass gap at strong coupling. We will now prove the existence of this mass
gap from the solution to equation \eqn{gdh4}. We note first that \eqn{gdh4} can be written
again as a Schr\"odinger equation. Indeed let
\be
\psi^{(2)}={\sinh^{3/2} z\ov \cosh^{1/2} z} \Psi(z)\ ,\qq \sinh z = {r_0\ov r}\ ,\quad
z\ge 0\ .
\ee
Then $\Psi$ satisfies the standard Schr\"odinger equation with potential
\ba
V(z)= (l+1)^2-{n^2-1/4\ov \cosh^2 z}   + {15/4+E_{m,l,n}\ov \sinh^2 z}\ ,
\label{pot1}
\ea
and eigenvalue $M^2R^4/r_0^2$. This belongs to the family of P\"oschl--Teller
potentials in quantum mechanics of type II. The potential decreases monotonically from
arbitrarily large positive values, where it behaves as in the conformal case in \eqn{pot1con},
to the constant $(l+1)^2$ as $z$ ranges from $0$ to $\infty$. Hence we expect a continuous
spectrum with a mass gap given by
\ba
M_{{\rm gap},l}= (l+1) {r_0\ov R^2} \ ,
\label{masgg}
\ea
which is degenerate according to \eqn{deg1}. In order to explicitly solve equation \eqn{gdh4} we
perform the change of variables
\be
\psi^{(2)}= \zeta^{-1/2+i \d/2} (1-\zeta)^{m+|n|/2+2} F(\zeta)\ ,\qq
\zeta ={r^2\ov r^2+r_0^2}={1\ov \cosh^2 z}\ ,\quad 0\le \zeta\le 1\ ,
\ee
where $\d$ is a constant to be determined. After some algebra we obtain a standard hypergeometric
equation for $F(\zeta)$, whose general solution leads to
\ba
\psi^{(2)}_{l,m,n}(r)  & = &
\cN \zeta^{-1/2} (1-\zeta)^{m+|n|/2+2} \left(e^{i\varphi} \zeta^{i\d/2}
{}_2F_1\left(a,a,1+i \d,\zeta\right)+c.c.\right)\ ,
\nonumber\\
&=& \cN \zeta^{-1/2} (1-\zeta)^{-m-|n|/2}\left(e^{i\varphi} \zeta^{i\d/2}
F(b,b,1+i+i \d,\zeta)+ c.c.\right)
\ea
where $\cN$ is an overall normalization constant and
\be
a={1\ov 2}(|n|+2 m +3+ i \d)\ ,\qq b=1+i\d -a\ ,\qq
\d=\sqrt{M^2 R^4/r_0^2-(l+1)^2}\ .
\ee
The phase $\varphi$ is computed by demanding that the solution is regular at $\zeta=1$
(equivalently at $r\to \infty$). We find that
\be
\varphi ={\pi\ov 2}+  {\rm Arg}\left(\G^2(a)\ov \G(1+i \d)\right)\ .
\ee
As in \cite{warn,BS}, where the $s$-wave case with $l=n=0$ was studied in detail,
 normalizability of the solution in the Dirac sense requires that
the parameter $\d$ is real and therefore the spectrum is continuous, with the
mass gap \eqn{masgg}.

%%%%%%%%%%%%%%%%%%%%%%%%%%%%%%%%%%%%%%%%%%%%%%%%%%%%%%%%%%%%%%%%%%%%%%%%
%%%%%%%%%%%%%%%%%%%%%%%%%%%%%%%%%%%%%%%%%%%%%%%%%%%%%%%%%%%%%%%%%%%%%%%%

\subsubsection{The sphere}

As in the previous cases eq. \eqn{gdh4} (with $r_0^2\to -r_0^2$)
can also be written as a Schr\"odinger equation. Now let
\ba
\psi^{(2)}={\sin^{3/2} z\ov \cos^{1/2} z} \Psi(z)\ ,\qq \sin z = {r_0\ov r}\ ,\quad
0\le  z\le \pi/2 \ .
\ea
Then $\Psi$ satisfies the standard Schr\"odinger equation with potential
\ba
V(z)= -(l+1)^2+{n^2-1/4\ov \cos^2 z}   + {15/4+E_{m,l,n}\ov \sin^2 z}\ ,
\label{pot2}
\ea
and eigenvalue $M^2R^4/r_0^2$. This belongs to the family of P\"oschl--Teller
potentials in quantum mechanics of type I. To find the
explicit solution we perform a change variables again through
\be
\psi^{(2)} = (1-\zeta)^{m+l/2+|n|/2} (1+\zeta)^{|n|/2} F(\zeta)\ ,
\qq \zeta=1-2 {r_0^2\ov r^2} \ ,\quad |\zeta|\le 1\ .
\ee
Then we obtain the standard Jacobi differential equation, with $\a=2 m +l+n-2$
and $\b=n$. The corresponding normalized solutions are
\ba
\psi^{(2)}_{k,m,l,n}(r) & = & C_{k,m,n}^2
\left(r_0\ov r\right)^{2 m+l+|n|+4} \left(1-{r_0^2\ov r^2}\right)^{|n|/2}
P_k^{(2 m+l+|n|+2,|n|)}(1-2r_0^2/r^2)\ ,
\nonumber\\
&&  k=0,1,\dots \ ,
\ea
where the normalization constant obeys
\be
C_{k,m,n}^2 = 2 (2 k+2 m+l+ 2|n|+3) {\G(k+1) \G(k +2m+l +2|n| +3)\ov \G(k+2 m+l+|n|+3) \G(k+|n|+1)}
{1\ov r_0^2}\ .
\ee
The mass eigenvalue is
\ba
M^2_{k,m,l,n}= 4(k+m+|n|+1)(k+m+|n|+l+2){r_0^2\ov R^4}\ .
\label{mmeig}
\ea
The measure that determined the normalization constant is $dr r$, with $r_0\le r < \infty$.
This mass eigenvalue is of the form $M^2=4 (j+1)(j+l+2)r_0^2/R^4$, with $j=k+m+|n|$.
Its degeneracy follows from a computation similar to that in \eqn{deg22},
\ba
d_{{\rm sphere},l,j}&=&d_{3,l} \sum_{m,k=0}^\infty  \sum_{n=-\infty}^\infty
\d_{k+m+|n|,j} = (l+1)^2 \sum_{m,p=0}^\infty \d_{m+p,j}\left[\sum_{k=0}^\infty
\sum_{n=-\infty}^\infty \d_{k+|n|,p}\right]
\nonumber\\
& = & (l+1)^2 \sum_{m,p=0}^\infty (2 p+1) \d_{m+p,j}
\label{fef2}
\\
& = & (l+1)^2(j+1)^2\ .
\nonumber
\ea

%%%%%%%%%%%%%%%%%%%%%%%%%%%%%%%%%%%%%%%%%%%%%%%%%%%%%%%%%%%%%%%%%%%%%%%%
%%%%%%%%%%%%%%%%%%%%%%%%%%%%%%%%%%%%%%%%%%%%%%%%%%%%%%%%%%%%%%%%%%%%%%%%

\subsection{The deformed case}

In this case we have to take into account the second term in the right
hand side of \eqn{jedhg}. In general this term destroys the $SO(4)$
spherical symmetry, unless we consider solutions that are independent of
the angles $\phi_{2,3}$ of $S^3$, as parametrized in \eqn{lpj2}.
We will then consider the Laplace operator within the ansatz
\ba
\Psi= {1\ov (2\pi)^{2}} e^{ik\cdot x}  \Psi_\perp(r,\th,\phi_1,\psi)\ ,
\label{ans2}
\ea
which is a restricted version of \eqn{ans1}. We obtain now an expression similar to
\eqn{hfdjew}, with an additional term inside the bracket. In particular,
\ba
\Box \Psi & \sim & {r\ov R^2 \sqrt{r^2+r_0^2\cos^2 \th}}
\Bigg[ \cdots +\hat \g^2 \cos^2\th \del^2_{\phi_1} \Bigg]\Psi_\perp \ ,
\label{hfdj2}
\ea
where the ellipsis denotes the terms inside the bracket in \eqn{hfdjew}. The measure
in the Hilbert space is given generally by $\sqrt{-G} e^{-2\Phi} G^{tt}$. This
is given, based on general grounds,\footnote{Since $e^{-2\Phi} \sqrt{G}$ is, in general,
invariant under T-duality and $G^{tt}$ is not affected by the T-dualities we performed.}
by the right hand side of \eqn{meas1}, but it can be
also verified by an explicit calculation.

Proceeding as in the undeformed case we make for the amplitude the
ansatz\footnote{If we make a general ansatz of the form
$\Psi_\perp\sim e^{i n_i \phi_i} \psi^{(1)}(\psi,\th) \psi^{(2)}(r)$,
where $n_i\in \mathbb{Z}$, the modification of the differential equation
depends on the matrix $\l_{ij}$ in \eqn{jedhg} via a term of the
form $-\g^2 \l_{ij}n_in_j$. For general integers $n_i$ this ansatz will
break the spherical $SO(4)$ symmetry of our background and there is no
longer factorization of the $\psi$ and $\th$ dependencies. However, if $n_2=n_3$
the spherical symmetry is preserved. Then using \eqn{zzzs} we may easily
show that the expressions below are still valid provided we replace $n^2$
by $ (n-n_2)^2$.}
\be
\Psi_\perp(r,\th,\phi_1,\psi) = {1\ov \sqrt{2\pi}} e^{i n\phi_1}
\Psi_{S^3}(\psi) \psi^{(1)}(\th)\psi^{(2)}(r)\ , \qq n\in \mathbb{Z} \ .
\ee
For $\Psi_{S^3}$ we obtain the same equation as in \eqn{sph3} but with no
$\phi_{2,3}$ for the solution. Hence, from \eqn{doo3} we have that
$l=0,2,\dots$ so that we make the replacement $l\to 2l$. Now the eigenvalues
are of the form $4l(l+1)$ and the normalized solution is given in
terms of Legendre polynomials as
\be
\Psi_{S^3,l} = 2 \left(l+\ha\right)^{1/2}\ P_l(\cos 2\psi)\ ,\qq l=0,1,\dots\ .
\ee
For the radial factor $\psi^{(2)}$ we obtain the same equations as in \eqn{gdh4}
with the replacement $l\to 2 l$. The only major modification is in the equation
for the angular part $\psi^{(1)}$ which now gets an additional term and becomes
\ba
{1\ov \sin\th \cos^3\th}{d\ov d\th} \sin\th \cos^3\th {d\psi^{(1)}\ov d\th} \!
+ \! \left[E-{n^2} \! \left( \! {1\ov \sin^2\th}+\hat\g^2 \cos^2\th \! \right) \!
 -{4 l(l+1)\ov \cos^2 \th} \right]\! \psi^{(1)} \! = 0  .
\label{gfh1}
\ea
This equation can also be written as a Schr\"odinger equation. Indeed let
\be
\psi^{(1)}={ \Psi(\th)\ov \sin^{1/2}\th\ \cos^{3/2}\th} \ .
\ee
Then $\Psi$ satisfies the standard Schr\"odinger equation with potential
\ba
V(\th)= -4+{n^2-1/4\ov \sin^2\th}   + {4l(l+1) +3/4\ov \cos^2 \th}\ +n^2\hat \g^2 \cos^2\th\
\label{pot13}
\ea
and eigenvalue $E$. When $n=0$ (or $n=n_2=n_3$, see footnote 13), the results
obtained for the undeformed case carry over unchanged. For $n\neq 0$ we could
not solve this equation by elementary means
for reasons related to the change in nature of the
singularity at infinity, as described in footnote 11. Nevertheless, we may resort to
perturbation theory which is valid for small values of the effective parameter $n^2 \hat \g^2$,
and to an asymptotic expansion for large values of the same parameter. We will return to
the generic case later.

%%%%%%%%%%%%%%%%%%%%%%%%%%%%%%%%%%%%%%%%%%%%%%%%%%%%%%%%%%%%%%%%%%%%%%%%
%%%%%%%%%%%%%%%%%%%%%%%%%%%%%%%%%%%%%%%%%%%%%%%%%%%%%%%%%%%%%%%%%%%%%%%%

\subsubsection{The case with $\hat \g^2 n^2 \ll 1 $}

In the limit $\hat \g^2 n^2 \ll 1 $ we can treat the last term in the potential
\eqn{pot13} as small. The corresponding shift in the energy eigenvalue can then be found
from conventional perturbation theory. Using the wavefunction \eqn{fgjw} we get
\ba
\d E_{m,l,n} & = & \ha n^2 \hat \g^2 B^2_{m,l,n}
\int_{-1}^{+1} dx (1-x)^{|n|} (1+x)^{l+2}\ [P_{m}^{|n|,l+1}(x)]^2\ > 0 \ .
\label{hgfg3}
\ea
For $l=0$ the integral can be computed. We found that
\be
\d E_{m,0,n}= 2^{4 + |n|} {(m+1)(m+|n|+1)\ov (2m +|n|+3)(2 m + |n|+1)} \ n^2 \hat \g^2 \ .
\ee
The corresponding change in the eigenfunctions $\psi^{(1)}$ is also computable
using perturbation theory, but it will not be presented here. The change in the values
of $E_{m,l,n}$ are affecting the differential equation for the radial factor
$\psi^{(2)}$ according to \eqn{gdh4}. It can be easily seen, for instance from
the form of the potentials for the effective Schr\"odinger problems \eqn{pot1con},
\eqn{pot1} and \eqn{pot2}, that the effect due to the shift $\d E_{m,l,n}$ can be
absorbed by an effective shift in the quantum number $m$. This can be computed by
varying \eqn{jhgf} with respect to $m$ and identifying the result with the perturbative
result \eqn{hgfg3}. In this way we find
\be
\d m =  {\d E_{m,l,n}\ov 4 (l+|n|+2 m + 2)}\ .
\ee
For the conformal and the disc cases the continuum mass spectrum for $M^2$
remains unchanged, except of course for the density of states which depends on the
wavefunction. For the sphere case the spectrum changes as
\be
\d M^2_{k,m,l,n}= {2 k +2 m + 2 |n| +l+3\ov 2 m+ |n|+2}\ \d E_{k,m,l,n} > 0  \ .
\ee

%%%%%%%%%%%%%%%%%%%%%%%%%%%%%%%%%%%%%%%%%%%%%%%%%%%%%%%%%%%%%%%%%%%%%%%%
%%%%%%%%%%%%%%%%%%%%%%%%%%%%%%%%%%%%%%%%%%%%%%%%%%%%%%%%%%%%%%%%%%%%%%%%

\subsubsection{The case with $\hat \g^2 n^2\gg 1 $}

In this case for generic values of the angle $\th$, i.e. not near $\th = 0, \pi/2$,
the potential goes to $+\infty$ and the wave function has the form of plane waves
of very high energy.
However, there are special states that are actually quasi-localized or even localized
near $\th = 0$ and $\th= \pi/2$, respectively.

\no
$\bullet $ {\underline {Region near $\th=0$}}: Let the change of variables
\be
\th = {z\ov n \hat \g}\ .
\ee
Then, within our limit the variable $z$ becomes non-compact with $z\ge 0$ and
the problem becomes equivalent to a Schr\"odinger problem with potential
\be
V(z)= {n^2 - 1/4\ov z^2} +1\ , \qq z\ge 0 \ ,
\ee
so that the spectrum is continuous with mass gap.
The solution is
\be
\psi^{(1)}\sim \sqrt{\bar E -1}\  J_{n}\left(\sqrt{\bar E -1}z\right)\ ,
\qq \bar E= {E\ov n^2 \hat \g^2} \ ,
\ee
with the measure induced by \eqn{meas1} being $dz z$. Note also that in our limit
$\psi^{(1)}\sim z^{-1/2}\Psi$. This solution is quasi-localized near $z=0$
(for $\th\sim {1\ov n\hat \g}$) in the sense that it dies off oscillating
with a power law behaviour away from it.

\no
$\bullet $ {\underline {Region near $\th=\pi/2$}}: Let the change of variables
\be
\th = {\pi \ov 2} -{z\ov \sqrt{n \hat \g}}\ .
\ee
Then, as before, within our limit the variable $z$ becomes non-compact with $z\ge 0$ and
the problem becomes equivalent to a Schr\"odinger problem with potential
\be
V(z)= {4 l(l+1) +3/4\ov z^2} +z^2\ , \qq z\ge 0 \
\ee
and rescaled energy parameter $\bar E = {E\ov n\hat \g}$.
The essential difference with the previous case is that the potential gets a term
corresponding to a harmonic oscillator.
The Schr\"odinger equation can be transformed into a confluent hypergeometric
equation. The solution regular for $z=0$ is then given by
\be
\psi^{(1)}\sim z^{2 l} e^{-z^2/2} F(l+1-\bar E/4,2 l+ 2,z^2)\ .
\ee
The measure induced by \eqn{meas1} is $dz z^3$. Note also that in our limit
$\psi^{(1)}\sim z^{-3/2}\Psi$.
For generic values of $\bar E$, the confluent hypergeometric function appearing in the
solution diverges exponentially as $e^{z^2}$, as $z\to \infty$.
In order to avoid this behaviour
at infinity we demand polynomial solutions
which implies a quantization for the eigenvalue $\bar E$,
as expected from the shape of the potential. Writing the result for the
original energy constant $E$ we have that
\be
E_{m,l,n} \simeq 4 (m+l+1) \ n\hat \g\ ,\quad m=0,1,\dots \qq {\rm as}\quad n\hat \g\gg 1\ ,
\ee
When $\hat \g\gg 1$ this formula is valid for $n\ge 1$.
Obviously, the solution is localized
near $z=0$ (for ${\pi\ov 2}-\th \sim  {1\ov \sqrt{n\hat \gamma}}$) in the sense that it
dies off exponentially away from it.

%%%%%%%%%%%%%%%%%%%%%%%%%%%%%%%%%%%%%%%%%%%%%%%%%%%%%%%%%%%%%%%%%%%%%%%%
%%%%%%%%%%%%%%%%%%%%%%%%%%%%%%%%%%%%%%%%%%%%%%%%%%%%%%%%%%%%%%%%%%%%%%%%

\subsection{Relation to the Heun equation and the Inozemtsev model}

We turn now to a general study of the equation \eqn{gfh1} for the angular
function $\psi^{(1)}(\th)$ in the deformed case. The substitution
\be
\psi^{(1)}= (1-z)^{|n|/2} z^{l} F(z)\ ,\qq z=\cos^2\th\ ,
\ee
gives the following equation for the function $F(z)$
\be
{d^2F\ov dz^2} + \left({2(l+1)\ov z} +{|n|+1\ov z-1}\right) {dF\ov dz} +
{n^2 \hat \gamma^2 z - E+(2 l+|n|)(2 l+|n|+4)\ov 4 z(z-1)} F= 0\ .
\ee
This is a second order ordinary differential equation with two regular
singular points at $z=0$ and at $z=1$, together with an irregular singularity
at $z=\infty$. Hence it is expected to correspond to a confluent form of the
Heun differential equation (see, for instance, \cite{Ronveaux}). The latter is the
standard form of a Fuchsian differential equation with four regular singularities.
Similar to the case of the confluent hypergeometric equation, under a confluence
process two of the singularities are made to coincide resulting into an irregular one.

There is an appealing relation of the Heun differential equation to the integrable
$BC_1$ Inozemtsev model. The $BC_1$ Inozemtsev system is a one-particle quantum
mechanical model in one dimension with potential
\ba
V=\sum_{i=0}^3 l_i (l_i+1) \wp(z+\om_i)\ ,
\label{finig}
\ea
where $\wp(z)$ is the Weierstrass elliptic function, which is doubly periodic
in the $z$ variable with half-periods $\om$ and $\om'$.
The Weierstrass elliptic function obeys
\be
(\wp^\prime)^2 = 4 (\wp-e_1)(\wp-e_2) (\wp-e_3)\ ,% \wp^2 -g_2 \wp - g_3\ ,
\ee
where the constants $e_i$'s  obey $\wp(\om)=e_1$,
$\wp(\om+\om')=e_2$ and $\wp(\om')=e_3$. In case they are in a straight line in the
complex plane, $e_2$ is assumed to lie between $e_1$ and $e_3$. In addition, they sum
up to 0. In the notation of \eqn{finig} $\om_0=0$, $\om_1=\om$, $\om_2=\om'$ and
$\om_3=\om+\om'$. This model was shown to be completely integrable in \cite{Inozemtsev}
and it belongs to the class of $BC_N$ integrable quantum N-particle systems with $B_N$
symmetry. It is a generalization of the $BC_N$ Calogero--Moser--Sutherland model.

The expression of the Heun equation in terms of elliptic functions was essentially known
by Darboux. The explicit relation of the $BC_1$ Inozemtsev model to the Heun equation can be
found, for instance, in \cite{Ronveaux,takemuraII,takemuraIII}. The potential \eqn{finig} was
also studied in \cite{verdier}, in the context of the theory of elliptic solitons. It was
shown that if the above coupling constants $l_i$, with $i=0,1,2,3$, are all integers, then the potential
\eqn{finig} has the finite gap property.

We will now consider the so-called trigonometric
limit of the Schr\"odinger equation corresponding to \eqn{finig}. In the standard version
of this limit we end up with the trigonometric P\"oschl--Teller potential, which in
our case is the potential \eqn{pot13}, but with $\hat \g=0$. Therefore, if such a limit
is taken, we do not capture the deformation. Here we will take instead a type of trigonometric
limit which can take into account the effect of the deformation.
In the mathematics literature this was considered before in various forms
(see \cite{oshima,takemura4}). In general, in the trigonometric
limit, we take the limit $e_2\to e_3$. Then one of the half-periods becomes
imaginary infinity, say $\om'\to i \infty$.
In order to investigate this limit it is better to consider the series representation of the
Weierstrass function in terms of powers of the parameter $q=e^{i \pi \tau}$, where $\tau$ is the
modular parameter defined as $\tau=\om_3/\om_1$. We take the half-periods to be
$\om_1=\ha$, $\om_2=-{\tau+1\ov 2}$ and $\om_3=-\om_1-\om_2={\tau\ov 2}$.
Then using standard expansions for the Weierstrass function and keeping only the
relevant to our discussion terms, we find
\ba
&& \wp(x)=-{\pi^2\ov 3}+ {\pi^2\ov \sin^2\pi x} + 16 \pi^2 \sin^2\pi x\ q^2 + {\cal O}(q^4)\ ,
\nonumber\\
&& \wp(x+\om_1)=-{\pi^2\ov 3}+ {\pi^2\ov \cos^2\pi x} + 16 \pi^2 \cos^2\pi x \ q^2 + {\cal O}(q^4)\ ,
\nonumber\\
&& \wp(x+\om_2)=-{\pi^2\ov 3}
+ 8\pi^2 \cos 2\pi x\ q + 8 \pi^2 (1-2 \cos 4\pi x)\ q^2 + {\cal O}(q^3)\ ,
\label{expwei}\\
&& \wp(x+\om_3)=-{\pi^2\ov 3}
- 8\pi^2 \cos 2\pi x\ q + 8 \pi^2 (1-2 \cos 4\pi x)\ q^2 + {\cal O}(q^3)\ .
\nonumber
\ea
Next we scale and shift the energy as
\be
E\to \pi^2\left(E-{1\ov 3}\sum_{i=0}^3l_i(l_i+1) \right)\
\ee
and let
\be
l_{2}={c\ov q} + \ha \left({d\ov c}-1\right) \ ,\qq
l_{3}=-{c\ov q} + \ha \left({d\ov c}-1\right) \ ,
\ee
where $c,d$ are real constants.
Then the limit $q\to 0$ exists even though the parameters $l_2$ and $l_3$ in the potential
both go to infinity.
If we define the angular variable $\th=-\pi/2 +\pi x$, taking values in
the range $\th\in [-\pi/2,\pi/2]$ we obtain a one-dimensional Schr\"odinger equation
with potential
\be
V(\th)= {l_0(l_0+1)\ov \cos^2\th } +  {l_1(l_1+1)\ov \sin^2\th }
+ 2 c^2 (1-2 \cos 4\th) - 16 d \cos 2\th\ .
\ee
We call this
the generalized trigonometric limit, in distinction
to the ordinary one for which $c=d=0$.
The potential \eqn{pot13} corresponds, up to an additive constant,
to the particular case $c=0$, $d=-{1\ov 32} n^2 \hat \g^2 $, $l_0=2l +\ha$
and $l_1=n-\ha$. Also the range of $\th $ is restricted to $\th\in [0,\pi/2]$
(this can be done by requiring vanishing eigenfunctions for $\th=0$).

The advantage of having a precise relation of \eqn{pot13} with the Heun equation and
the $BC_1$ one-particle quantum Inozemtsev integrable model is the fact that techniques
have been developed based on the Bethe ansatz in order to determine the eigenvalue
problem for the general potential \eqn{finig}. In particular, in \cite{takemuraI} an
explicit solution was presented. However, it still involves satisfying $l+4$, with
$l=\sum_{i=0}^3l_i$, complicated relations involving Theta-functions for $l$ auxiliary
parameters. Solving this problem in general seems impossible. In \cite{takemuraI} it
was shown how to solve this problem in the ordinary trigonometric limit employing the Bethe
ansatz. The result reduces to Jacobi polynomials as in our case, in a presumably
equivalent solution. We expect that progress can be made using the Bethe ansatz method
even in the generalized trigonometric limit.

%%%%%%%%%%%%%%%%%%%%%%%%%%%%%%%%%%%%%%%%%%%%%%%%%%%%%%%%%%%%%%%%%%%%%%
%%%%%%%%%%%%%%%%%%%%%%%%%%%%%%%%%%%%%%%%%%%%%%%%%%%%%%%%%%%%%%%%%%%%%%

\section{Conclusions}

In this paper we have constructed type-IIB supergravity duals to the
Coulomb branch of a class of exactly marginal deformations of ${\cal N}=4$ supersymmetric
Yang--Mills theory. The solutions can be derived by applying a sequence of T-dualities and
coordinate shifts to multicenter supergravity backgrounds, similarly to the conformal
case \cite{LM}. On the gravity side of the AdS/CFT correspondence the marginal perturbation leads
to a deformation of the space transversal to the worldvolume of the branes. We have
probed the geometry of the marginal deformation by computing the expectation value of the
Wilson loop operator from a string extending along a great circle of this deformed
transversal space. The cases that we have considered in detail correspond to a
continuous distribution of D3-branes on a disc or on the surface of a three-dimensional
sphere, and preserve an $SO(4) \times SO(2)$ global symmetry group. In the conformal
limit, where the Higgs vev remains small, we observed the usual Coulomb behaviour
for the quark-antiquark potential. The background corresponding to a distribution
of branes on a disc is quite rich, and contains several regimes according to the
relation between the various parameters. In particular we have found situations where
the quark-antiquark interaction is completely screened as in \cite{BS},
or where a confining behaviour arises, with a linear or logarithmic potential according
to the ratio of the quark-antiquark separation distance to the Higgs vev scale. It would be
of great interest to explore this novel behaviour on the field theory side of the correspondence
and in particular to explore the origin of this logarithmic dependence.

We have also described the spectra of massless excitations by solving the Laplace
equation in the deformed $SO(4) \times SO(2)$ background. In order to solve the corresponding
differential equations we have transformed them into Schr\"odinger equations with a
potential determined by the geometry of the supergravity background. We have performed
a detailed analysis of the equation in the ${\cal N}=4$ limit, with no deformation,
for the conformal, the disc and the sphere distributions. It is quite remarkable that
it is possible to explicitly solve this problem beyond the $s$-wave approximation
that has been considered before \cite{warn,BS}. It will be interesting to extend the
analysis to the full class of models corresponding to the gravity-scalar sector (with scalars
in the coset $SO(6,\IR)/SO(6)$) of five-dimensional gauged supergravity that admit a
classification via a correspondence with algebraic curves \cite{Basfe2}. In the deformed
background we have relied on perturbation theory to find a solution for small values of the
deformation parameter, or on an asymptotic expansion in the opposite limit. Generic
values of the deformation lead to a confluent form of the Heun differential equation,
which is known to be related to the Inozemtsev integrable system. This relation provides
a tool to find solutions to the differential equation through the Bethe ansatz method.
In this article we have introduced a modification of the usual trigonometric
limit to give account of the deformation parameter. A complete attempt to investigate
the spectral problem in the generalized trigonometric limit that we have performed is definitely
worth the effort.

%%%%%%%%%%%%%%%%%%%%%%%%%%%%%%%%%%%%%%%%%%%%%%%%%%%%%%%%%%%%%%%%%%%%%%
%%%%%%%%%%%%%%%%%%%%%%%%%%%%%%%%%%%%%%%%%%%%%%%%%%%%%%%%%%%%%%%%%%%%%%

%\newpage

\vskip .4in

\centerline{ \bf Acknowledgments}

\no
K.~S. thanks N.~Brambilla for a useful correspondence.
K.~S. and D.~Z. acknowledge the financial support provided through the European Community's
program ``Constituents, Fundamental Forces and Symmetries of the Universe'' with contract MRTN-CT-2004-005104,
the INTAS contract 03-51-6346 ``Strings, branes and higher-spin gauge fields'', as well as
the Greek Ministry of Education programs $\rm \P Y\Th A\G OPA\S$ with contract 89194 and
$\rm E\Pi A N$ with code-number B.545. K.~S. also thanks CERN and the University of Neuch\^atel
for hospitality and financial support during part of this research.
In addition, D.~Z. acknowledges the financial support provided through the Research Committee
of the University of Patras for a ``K.~Karatheodory'' fellowship under contract number 3022.

\no
Parts of this research were presented by K.~S. in the RTN
meeting ``Constituents, Fundamental Forces and Symmetries of the Universe'',
in Corfu, Greece, 20-26 September 2005.

%%%%%%%%%%%%%%%%%%%%%%%%%%%%%%%%%%%%%%%%%%%%%%%%%%%%%%%%%%%%%%%%%%%%%
%%%%%%%%%%%%%%%%%%%%%%%%%%%%%%%%%%%%%%%%%%%%%%%%%%%%%%%%%%%%%%%%%%%%%

\end{document}